\documentclass[fleqn,usenatbib]{mnras}

\usepackage{newtxtext,newtxmath}

\usepackage[T1]{fontenc}
\usepackage{ae,aecompl}

\usepackage{graphicx}	
\usepackage{amsmath}	
\usepackage{gensymb}

\title[GW Ori circumtriple disc]{GW Ori: circumtriple rings and planets}

\author[Smallwood et al.]{Jeremy L. Smallwood,$^1$\thanks{E-mail: Smallj2@unlv.nevada.edu}
Rebecca Nealon$^{2,3,4}$,
Cheng Chen$^1$,
Rebecca G. Martin$^1$,
\newauthor
Jiaqing Bi$^5$,
Ruobing Dong$^5$ and
Christophe Pinte$^{6,7}$
\\
$^1$Department of Physics and Astronomy, University of Nevada, Las Vegas, 4505 South Maryland Parkway, Las Vegas, NV 89154, USA\\
$^2$Department of Physics and Astronomy, University of Leicester, University Road, Leicester LEI 7RH, UK\\
$^3$Centre for Exoplanets and Habitability, University of Warwick, Coventry CV4 7AL, UK\\
$^4$Department of Physics, University of Warwick, Coventry CV4 7AL, UK\\
$^5$Department of Physics and Astronomy, University of Victoria, Victoria, BC V8P 1A1, Canada\\
$^6$School of Physics and Astronomy, Monash University, Clayton Vic 3800, Australia\\
$^7$Univ. Grenoble Alpes, CNRS, IPAG, F-38000 Grenoble, France
}

\date{Accepted XXX. Received YYY; in original form ZZZ}

\pubyear{2020}

\begin{document}
\label{firstpage}
\pagerange{\pageref{firstpage}--\pageref{lastpage}}
\maketitle

\begin{abstract}
GW Ori is a hierarchical triple star system with a  misaligned circumtriple protoplanetary disc. Recent ALMA observations have identified three dust rings with a prominent gap at $100\, \rm au$ and misalignments between each of the rings. A break in the gas disc may be driven either by the torque from the triple star system or a planet that is massive enough to carve a gap in the disc. Once the disc is broken, the rings nodally precess on different timescales and become misaligned. We investigate the origins of the dust rings by means of $N$--body integrations and 3-dimensional hydrodynamic simulations. We find that for observationally-motivated parameters of protoplanetary discs, the disc does not break due to the torque from the star system. We suggest that the presence of a massive planet (or planets) in the disc  separates the inner and outer disc.  We conclude that the disc breaking in GW Ori is likely caused by undetected planets --- the first planet(s) in a circumtriple orbit.
\end{abstract}




\section{Introduction}

Observations show that the majority of stars form in relatively dense regions within stellar clusters, naturally leading to  multi-star systems. It is estimated that more than $40\%$-$50\%$ of stars are in a binary pair, while about $20\%$ are observed in triple or higher-order systems \citep{Duquennoy1991,Raghaven2010,Tokovinin2014a,Tokovinin2014b}. To date, there have been planets found in 32 triple star systems \citep{Busetti2018,Fragione2019}.
However, as yet no planet in a circumtriple orbit has been discovered.

GW Ori is one such hierarchical triple star system, with a spectroscopic binary (A,B) at a separation of about $1\, \rm au$ and a tertiary stellar companion (C) at a projected distance of about $8\, \rm au$ \citep{Mathieu1991,Berger2011,Czekala2017,Kraus2020}.  The system is at a distance of $408\pm 10\, \rm parsec$ \citep{GiaCollaboration2020} and has an age of  $1.0\pm 0.1\, \rm Myr$ \citep{Calvet2004}. The derived triple star parameters from \cite{Czekala2017} are shown in Table~\ref{table::binaryparams}. Stars A and B have masses of $2.8\, \rm M_{\odot}$ and $1.7\, \rm M_{\odot}$, respectively, while star C has a mass of $1.2\, \rm M_{\odot}$ and the eccentricity of the outer companion is estimated to be $0.22$ \citep{Czekala2017}.  The system has a circumtriple protoplanetary disc in which the dust is observed to extend to about $400\, \rm au$, while the gas extends farther to about $1300\, \rm au$ \citep{Fang2017}. 

\cite{Bi2020} presented Atacama Large Millimeter/submillimeter Array (ALMA) observations of the dust continuum and molecular gas emission of the circumtriple disc of GW Ori.  They identified three dust rings in the circumtriple disc at radii of about $46$, $188$, and $338\, \rm au$ from the triple star. The structure of the rings are outlined in Table~\ref{table::rings}.
The radial width of dust ring 3  is the widest ring ever found in a protoplanetary disc. The  dust masses of the  rings 1, 2, and 3  were estimated to be $74\pm 8$, $168\pm19$, and $245\pm 28\, \rm M_{\oplus}$ \citep{Bi2020}. Misalignments are present between each individual dust ring and triple star orbital plane from visibility modelling of dust continuum and CO kinematics. The inclination and the longitude of ascending node of the AB-C binary orbit was found to be $150\pm 7\degree$ and $28\pm 9\degree$, respectively \citep{Berger2011}. Assuming that the entire disc has the same clockwise on-sky projected orbital direction, \cite{Bi2020} found that rings 1, 2, and 3  were misaligned by $11\pm 6\degree$, $35\pm 5 \degree$ and $40\pm 5\degree$, respectively, relative to the orbital plane. Therefore, their results suggested that  ring 1 and the AB-C binary plane are close to being coplanar while the  rings 2 and 3 are misaligned but close to coplanar with each other. 

A potential explanation for the large misalignment between the inner and middle rings is the `disc breaking' phenomenon, where the torque from the misaligned stars overcomes the viscous stresses and pressure holding the disc together and separates the disc into distinct planes \citep[e.g.][]{Nixon2011,Nixon2012b,Facchini2013,Nixon2013a}. However, \cite{Bi2020} ran SPH simulations modelling the disc with an inner, misaligned binary (that approximates the triple star evolution) and found that the disc does not break only due to the triple star torque. They modelled the disc with an aspect ratio $H/r = 0.05$. For this case, the disc did undergo strong warping at the beginning of the simulation but later relaxed to a steady state resembling a flat disc. They conclude that the gaps in the dust cannot solely be caused by the gravitational torque of the system and thus there must be another cause for the gaps.

 \begin{table}
	\centering
	\caption{The location of the dust rings in the GW Ori circumtriple disc. The values were obtained by MCMC fitting from \citet{Bi2020}. The second column gives the inner edge of the ring. The third column denotes the centre location of the ring, and the fourth column represents the outer edge of the ring.}
	\begin{tabular}{cccc} 
		\hline
	    Ring & Inner Radius & Centre Radius & Outer Radius\\
         & (au)  & (au) & (au) \\
		\hline
        \hline
		1 & $36.9804^{+0.134}_{-0.152}$  & $46.5516^{+ 0.05}_{-0.048}$ & $56.1228^{+0.134}_{-0.152}$\\
		& & & \\
        2 & $153.9865^{+0.2534}_{-0.8105}$ & $188.2827^{+0.1065}_{-0.1037}$ & $222.5790^{+0.2534}_{-0.2567}$ \\
        & & & \\
        3 & $270.5542^{+0.7514}_{-0.8105}$ & $337.2438^{+0.3735}_{-0.3602}$ & $403.9334^{+0.7514}_{-0.8105}$ \\
        \hline
	\end{tabular}
    \label{table::rings}
\end{table}

 \begin{table*}
	\centering
	\caption{The orbital elements of the GW Ori triple system from {\protect\cite{Czekala2017}} and {\protect\cite{Kraus2020}}.}
	\begin{tabular}{l*4c} 
		\hline
	     & \multicolumn{2}{c}{\cite{Czekala2017}} & \multicolumn{2}{c}{\cite{Kraus2020}} \\
	    Parameter & Orbit $A$ -- $B$ & Orbit ($AB$) -- $C$ & Orbit $A$ -- $B$ & Orbit ($AB$) -- $C$ \\
	    \hline
        \hline
        Period $P$ (days) & $241.50 \pm 0.05$ & $4246\pm 66$ & $241.619 \pm 0.05$ & $4216.8\pm4.6$ \\
        Semi-major axis $a$ (au) & $1.25\pm 0.05$  & $9.19\pm 0.32$ & $1.20\pm 0.04$   &  $8.89\pm 0.04$ \\
        Eccentricity $e$ & $0.13 \pm 0.02$ & $0.22 \pm 0.09$ &  $0.069\pm 0.009$   & $0.379\pm 0.003$ \\
        inclination $i$ (\degree) & $157\pm1$  & $150\pm 7$ & $156 \pm 1$    &  $149.6 \pm 0.7$ \\
        Longitude of the periastron  $\omega^{\rm a}$ (\degree) & $17\pm7$  & $130\pm21$ &  $1 \pm 7$   &  $105 \pm 1$ \\
        Longitude of the ascending node $\Omega^{\rm a}$ (\degree) & $263\pm13$  & $282\pm9$  &  $258.2 \pm 1.3$   & $230.9 \pm 1.1$  \\
        Total mass $M_{\rm tot}$ ($\rm M_{\odot}$) &  $4.48^{+0.42}_{-0.36}$  & $5.63^{+0.58}_{-0.43}$  & $3.90 \pm 0.40$  & $5.26 \pm 0.22$\\
        Star A mass $M_{\rm A}$ ($\rm M_{\odot}$)  &  \multicolumn{2}{c}{$ 2.80^{+0.36}_{-0.31}$}  & \multicolumn{2}{c}{$2.47 \pm 0.33$}  \\
        Star B mass $M_{\rm B}$ ($\rm M_{\odot}$)  &  \multicolumn{2}{c}{$ 1.68^{+0.21}_{-0.18}$}    & \multicolumn{2}{c}{$1.43 \pm 0.18$}  \\
        Star C mass $M_{\rm C}$ ($\rm M_{\odot}$)  &  \multicolumn{2}{c}{$ 1.15^{+0.40}_{-0.23}$}     & \multicolumn{2}{c}{$1.36 \pm 0.28$} \\
		\hline
	\end{tabular}
    \label{table::binaryparams}
\end{table*}

Recently, \cite{Kraus2020} presented higher resolution ALMA dust observations of GW Ori.  They updated the triple star parameters which are shown in Table~\ref{table::binaryparams}. They found that $M_{\rm A} = 2.47 \pm 0.33\, \rm M_{\odot}$, $M_{\rm B} = 1.43 \pm 0.18\, \rm M_{\odot} $, and $M_{\rm C} = 1.36 \pm 0.28 \, \rm M_{\odot}$ and the eccentricity of the outer most companion was increased to $e_{\rm AB-C}=0.379$. They found three distinct coherent dust rings similar to \cite{Bi2020}. The eccentricity of  ring 1 is found to be $\sim 0.2$. Moreover, \cite{Kraus2020} ran SPH simulations to investigate the origin of the substructures located in the circumtriple disc.  They found that the torque from the triple system could effectively break the disc, finding a relative misalignment between  rings 1 and 2 that was consistent with the observations. 


The major focus of this work is to investigate the dynamical origin of the dust rings that are present in the GW Ori inclined circumtriple disc. In Section~\ref{sec::three-body}, we  begin by considering the behaviour of test particle orbits around the triple star versus a binary. In Section~\ref{sec::setup}, we describe our numerical setup  for the hydrodynamical simulations and in Section~\ref{sec::results}, the results. We discuss our findings in the context of the observations of GW Ori in Section~\ref{sec::discussion}, and summarise in Section~\ref{sec::conclusions}.

 \begin{figure*}
 \centering
    \includegraphics[width=\columnwidth]{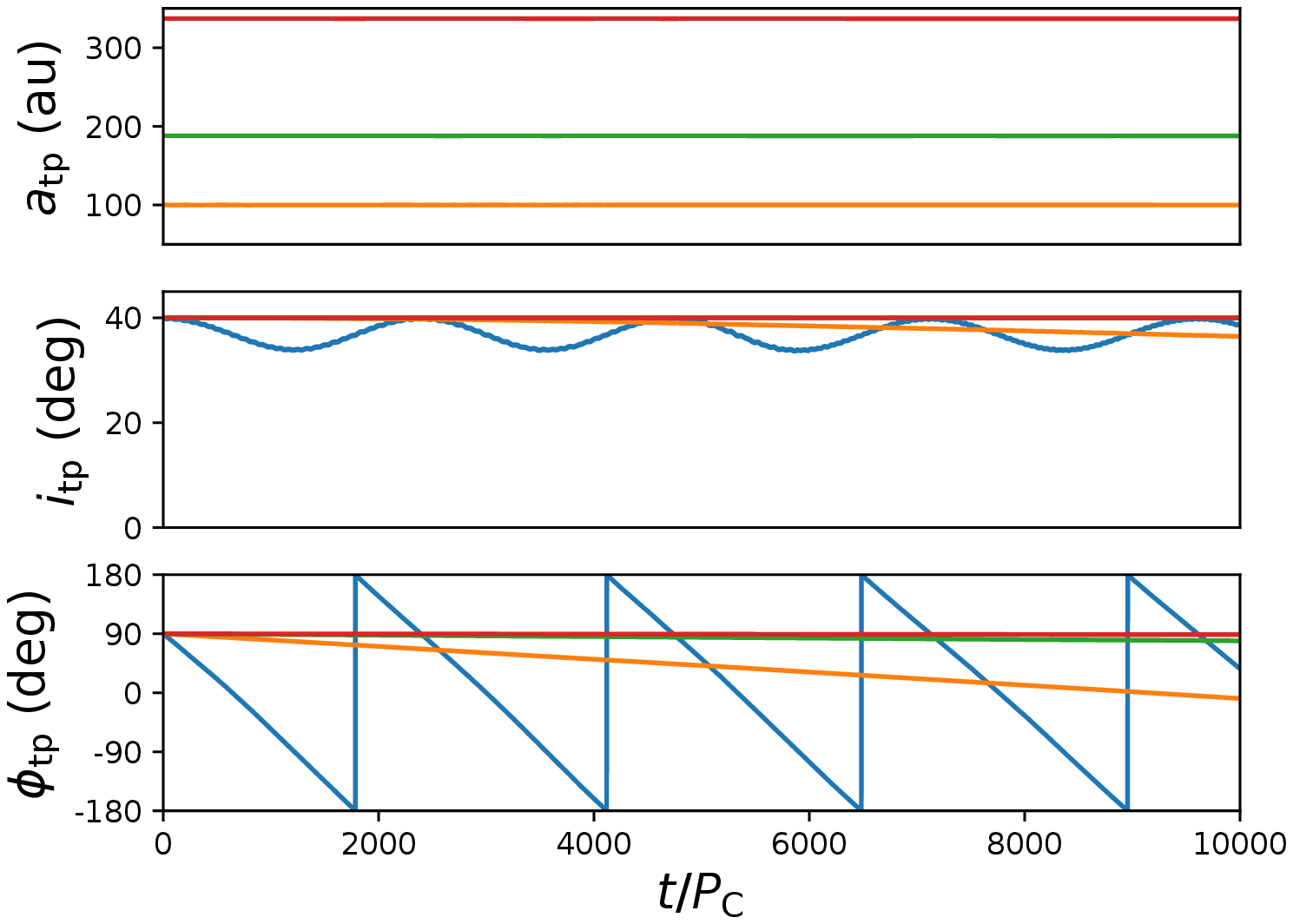}
    \includegraphics[width=\columnwidth]{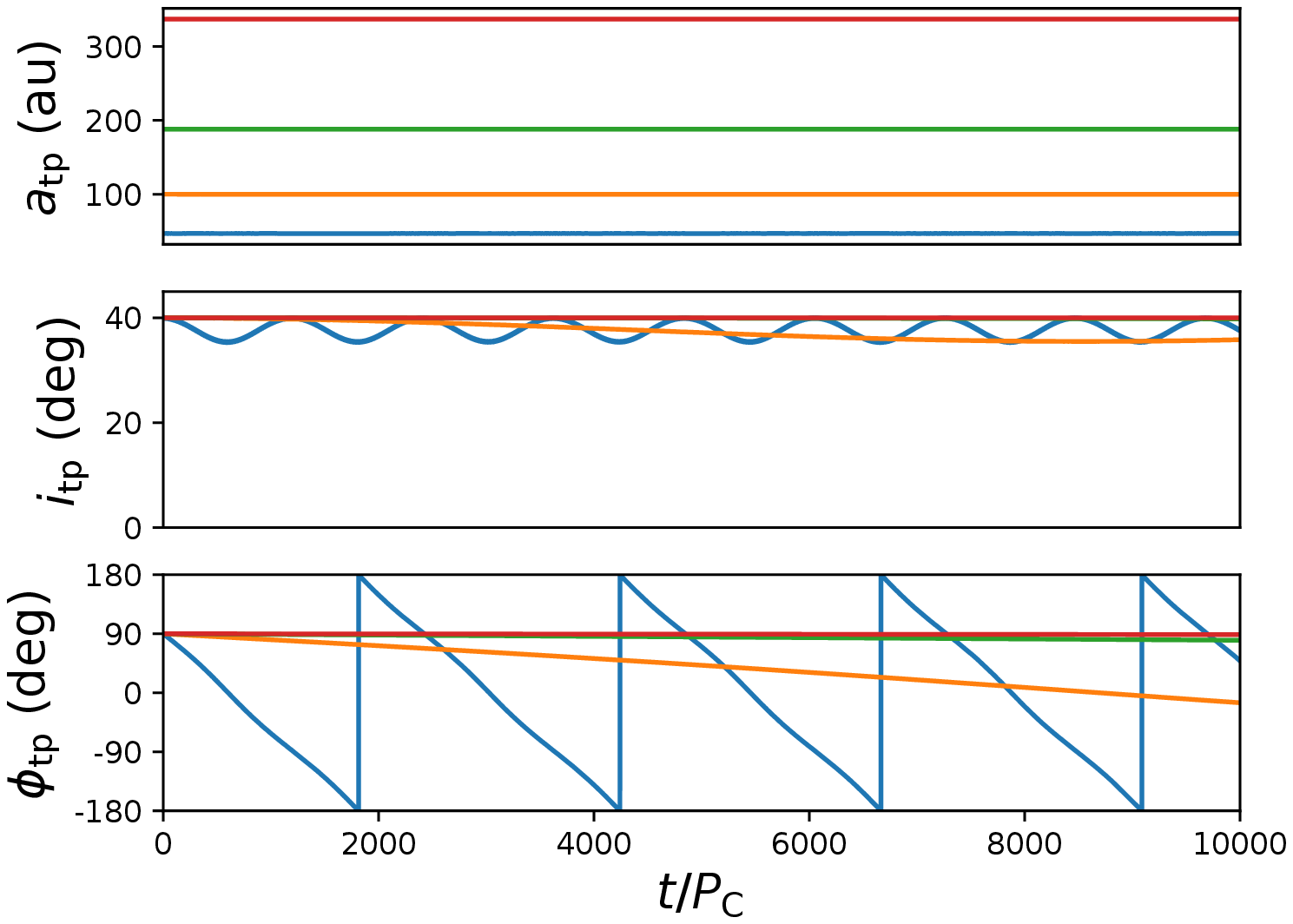}
    \caption{ Test particle evolution around the GW Ori triple star system (left panel) and around the approximated binary star system (right panel). We show the semi-major axis $a_{\rm tp}$, inclination $i_{\rm tp}$, and longitude of the ascending node $\phi_{\rm tp}$ of the test particles as a function of time. The stellar parameter set used is from {\protect\cite{Kraus2020}}. We show four different test particles that begin with an initial semi-major axis of $47$ (blue), $188$ (green), $337$ (red), and $100\, \rm au$ (yellow). The first three separations denote the center radius of the three observed dust rings, and the latter represents the center of the observed inner gap.}
\label{fig::rebound_tp}
\end{figure*}
 
\section{Four-Body Dynamics}
\label{sec::three-body}


In this section, we simulate the GW Ori system using the {\sc whfast} integrator which is a second order symplectic Wisdom Holman integrator with 11th order symplectic correctors in the {\sc n}-body simulation package, {\sc rebound} \citep{Rein2015}.  We explore the effect of modelling the close spectroscopic binary (stars A and B) as a single star, making the GW Ori system a binary (stars AB and C). We consider both a triple system (stars A, B, and C) and binary system (stars AB and C). We solve the gravitational equations for the three (two) bodies in the frame of centre-of-mass of the three (two) body system.  The parameters used for both the triple system and the binary system  are given in Table~\ref{table::binaryparams}. 

 Test particle orbits around binary and higher-order systems can show qualitatively similar behaviour to rings in a gas disc \cite[e.g.,][]{Doolin2011,Martin2014,Smallwood2019a,Chen2019b}.  Thus, in order to investigate the effect of simplifying the three star configuration to a binary system, we simulate the evolution of test particles around each stellar geometry in Fig.~\ref{fig::rebound_tp}. The stars' orbits are similar regardless of using three stars or a binary, and additionally, the orbits are stable over a long timescale.  We select four different semi-major axes of the particles of $47$, $188$, $337$, and $100\, \rm au$. The first three separations denote the center radius of the three observed dust rings, and the latter represents the center of the observed inner gap. The tilt of the test particle, $i_{\rm tp}$, is  measured from the $z$-axis and the phase angle of the test particle, $\phi_{\rm tp}$, is measured using the angular momentum unit vector of the test particle in the  $y$ and $x$ directions. 
For the binary case, the test particles nodally precess about the angular momentum vector of the binary and show small tilt oscillations as a result of the eccentricity of the AB-C binary \citep[e.g.][]{Smallwood2019a}.   Test particles around the triple star system also undergo small tilt oscillations that are driven by the precession of the triple stars.

The timescale for these tilt oscillations increases with test particle separation. The particle dynamics are not significantly affected by the presence of the AB binary (the left panel includes the triple star system and the right panel the binary approximation). The amplitude of the tilt oscillations are slightly greater around the triple star, by only a few degrees, than around the binary star approximation. The frequency of the tilt oscillations occur on a faster timescale around a binary than around the triple star. However, in both cases, the amplitude of the tilt oscillations are small. 

 The nodal precession timescale for the test particles is very similar around the triple star compared to the binary star approximation.
Since the nodal phase angle is changing on a much faster timescale than the tilt, we expect that in a disc simulation, the disc will remain relatively unwarped (constant inclination with radius) but it may become twisted (variation of nodal phase angle with radius). Disc breaking as a result of the binary is therefore likely driven by the precession. Since the precession is very similar in the binary and triple star cases, we expect a similar amount of warp/twist to the disc around two stars as around a triple star system.
In Section~\ref{sec::results} we first use hydrodynamical gas disc simulations around the binary and triple star system to confirm this.

 \begin{table*}
	\centering
	\caption{ The set-up of the SPH simulations that lists the  number of sink particles, inner circumbinary (circumtriple) disc radius $r_{\rm in}$, outer circumbinary  (circumtriple) disc radius $r_{\rm out}$, initial circumbinary (circumtriple) disc tilt $i_{0}$, disc aspect ratio $H/r$, the input \citet{shakura1973} $\alpha$-viscosity parameter, surface density power-law index $p$, the mean shelled-averaged smoothing length per scale height $\langle h \rangle/H$, mass of the planet $M_{\rm p}$, initial inclination of the planet $i_{0,\rm p}$, and the specific set of binary parameters.  The bold highlights the defining parameter for each simulation.}
	\begin{tabular}{cccccccccccc} 
		\hline
	    Simulation  & \# of Stars & $r_{\rm in}$ & $r_{\rm out}$ & $i_{0}$ & $H/r$ & $\alpha$ & $p$ & $(\langle h \rangle/H)_{\rm mean}$ & $M_{\rm p}$ & $i_{\rm 0,p}$ & Stellar Parameters \\
         & & (au)  & (au) &  $(\degree)$ & & & &  & ($M_{\rm J}$)  & $(\degree)$  &  \\
		\hline
        \hline
        run0 & $\bf 3$ & $40$ & $200$  & $38$  & $0.05$ & $0.01$  & $0.5$ & $0.30$    & -- & --  &  \cite{Kraus2020}  \\
        \hline
		run1 & $2$ & $40$ & $400$  & $38$  & $\bf 0.05$ & $0.01$  & $1.5$ & $0.32$    & -- & --  & \cite{Czekala2017} \\
        run2 & $2$ & $40$  & $400$  & $40$  & $0.05$ & $0.01$  & $1.5$ & $0.32$ & -- & --  & {\bf \cite{Kraus2020}} \\
        run3 & $2$ & $40$  & $400$  & $40$  & $\bf 0.1$ & $0.01$  & $1.5$ & $0.20$  & -- & --  & \cite{Czekala2017}  \\
        \hline
        run4 & $2$ & $40$  & $400$  & $40$  & $\bf 0.05$ & $0.01$  & $1.5$ & $0.32$   & $1$ & $40$  & \cite{Czekala2017}  \\
        run5 & $2$ & $40$  & $400$  & $40$  & $\bf 0.1$ & $0.01$  & $1.5$ & $0.20$  & $1$ & $40$  & \cite{Czekala2017}  \\
        \hline
        run6& $2$  & $\bf 40$  & $200$  & $38$  & $0.05$ & $0.01$  & $0.5$ & $0.30$   & -- & --  & \cite{Kraus2020}  \\
        run7& $2$  & $40$  & $200$  & $38$  & $0.05$ & $\bf 0.05$  & $0.5$ & $0.30$    & -- & --  & \cite{Kraus2020}  \\
        run8& $2$  & $40$  & $200$  & $38$  & $0.05$ & $\bf 0.1$  & $0.5$ & $0.30$      & -- & --  & \cite{Kraus2020}  \\
        run9& $2$  & $\bf 30$  & $200$  & $38$  & $0.05$ & $0.1$  & $0.5$ & $0.32$      & -- & --  & \cite{Kraus2020}  \\
        run10& $2$  & $\bf 20$  & $200$  & $38$  & $0.05$ & $0.01$  & $0.5$ & $0.34$  & -- & --  & \cite{Kraus2020}  \\
        \hline
	\end{tabular}
    \label{table::setup}
\end{table*}

\section{Numerical Methods}
\label{sec::setup}

For our numerical simulations we use the 3-dimensional smoothed particle hydrodynamics \cite[SPH; e.g.,][]{Price2012} code {\sc phantom} \citep{Price2018}.  We simulate an accretion disc around a binary and triple star system. {\sc phantom} has been well tested and used to model misaligned accretion discs in binary systems \citep[e.g.][]{Nixon2013a,Martin2014,Franchini2019}.  A misaligned disc feels a gravitational torque exerted by a binary or triple star. This causes the disc to undergo differential precession which can lead to disc warping, `breaking' \citep{Nixon2013a,Facchini2013} or `tearing' \citep{Nixon2012a}. Dissipation within the misaligned disc causes the differential precession to generate a warp which will evolve in the  diffusive regime or the bending wave regime, depending on the disc thickness and viscosity \citep{Paploizou1983,papaloizou1995,Ogilvie1999}. In the bending wave regime the disc aspect ratio is larger than the  viscosity coefficient ($H/r \gtrsim \alpha$),  and the warp induced in the disc by the binary torque propagates as a pressure wave with speed $c_{\rm s}/2$ \citep{Paploizou1983,papaloizou1995}. For the diffusive regime the disc aspect ratio is less than the viscosity ($H/r\lesssim \alpha$) with a diffusion coefficient inversely proportional to the disc viscosity. Most of our simulations are in the bending wave regime. Although we do consider some simulations in the diffusive regime for demonstrative purposes, we note that protoplanetary discs are expected to be in the bending-wave regime \citep{Hartmann1998}.  We discuss this further in Section~\ref{visc}.

In total we conduct  11 simulations that are summarised in Table~\ref{table::setup} and  we present additional information of our simulation parameters in Table~\ref{table::A1}.  First, we compare the disc structure around three stars versus two stars. Second, we compare the two different sets of stellar parameters from \cite{Czekala2017} and \cite{Kraus2020}. We then compare various disc aspect ratios and \cite{shakura1973} $\alpha$ values.  Furthermore, we also run two simulations that include  a planet  that is initially coplanar to the disc. The simulations without a planet are simulated for $3000\,\rm P_{\rm orb}$ and simulations with a planet are simulated for $230\, \rm P_{\rm planet}$, where $P_{\rm orb}$ is the binary orbital period and $P_{\rm planet}$ is the planet orbital period. For a $1\, \rm M_{\rm J}$ planet at $100\, \rm au$, $1 P_{\rm planet} \approx 36 P_{\rm orb}$,  so therefore the planet simulations ran for $\sim 8000\, \rm P_{\rm orb}$. The simulations without a planet have reached a steady-state within $3000\, \rm P_{orb}$. \cite{Kraus2020} ran their simulation for a much shorter time of $\sim 860\, \rm P_{\rm orb}$. 

We additionally generate synthetic CO maps for two of our simulations for comparison with the ALMA $^{12}$CO $J=2-1$ first moment map from \citet{Bi2020}. For this we use the Monte Carlo radiative transfer code \textsc{MCFOST} \citep{Pinte2006,Pinte2009}. \textsc{MCFOST} is particularly well suited for use with a particle based numerical method because it uses a Voronoi mesh (rather than a cylindrical or spherical grid) to generate a grid from the particles. Because the mesh follows the particle distribution it does not require any interpolation.
 
\subsection{Triple star parameters}
We setup one hydrodynamical simulation with three stars that are modelled as sink particles. The triple star parameters that we adopt are from \citet{Kraus2020}, which is summarised in Table~\ref{table::binaryparams}. The second column in Table~\ref{table::setup} details the stellar setup (either a binary or triple star system) used in each simulation. The inner binary and tertiary star begin at apastron. The  components of the tight A-B binary have an accretion radius of $R_{\rm acc}=0.5\, \rm au$, while the tertiary companion has $R_{\rm acc}=2.3\, \rm au$. Particles within the hard accretion radius are considered accreted and their mass and angular momentum are added to the respected star. 
 
\subsection{Binary parameters}
In Section~\ref{sec::three-body}, we showed that we can model the GW Ori hierarchical triple system as a AB-C binary. To model the binary we use the two sets of binary orbital parameters measured by \citet{Czekala2017} and \citet{Kraus2020}. Table~\ref{table::binaryparams}  compares the binary parameters from each study.  The last column in Table~\ref{table::setup} details which simulation uses which set of binary parameters. The binary begins at apastron and the accretion radius of each binary component is $R_{\rm acc}=4\, \rm au$, regardless of which binary parameters are adopted. Particles within this radius are accreted and their mass and angular momentum are added to the star.

\subsection{Disc setup}


Each simulation consists of $10^6$ equal mass Lagrangian particles initially distributed from the inner disc radius, $r_{\rm in}$, to the outer disc radius, $r_{\rm out}$. We consider two values of the inner radius, $20\, \rm au$ and $40\, \rm au$. The latter is farther out than the radius where the tidal torque truncates the disc \citep{Artymowicz1994} -- although we note that for a misaligned disc the tidal torque produced by the binary is much weaker, allowing the disc to survive closer to the binary \cite[e.g.,][]{Lubow2015,Miranda2015,Nixon2015,Lubow2018}. The observed outer radius of the gas disc is $\sim 1300\, \rm au$ \citep{Bi2020}, which means that the majority of the angular momentum is in the outer regions of the disc. In our simulations we truncate the outer radius to be $r_{\rm out}=200\, \rm au$ or $400\, \rm au$ in order to speed up computational time  and increase the resolution.  We note that this truncated outer radius preserves the angular momentum balance (i.e., the outer disc still holds most of the angular momentum).

The total disc mass is set to $0.1 \, \rm M_{\odot}$ assuming a dust to gas ratio of $0.01$. The value of the disc mass comes from the observations of the dust mass \cite[e.g.,][]{Bi2020}.  \cite{Kraus2020} estimated a lower disc mass because they did not recover the total flux due to missing short baselines in their ALMA observations. We ignore the effect of self-gravity since it has no effect on the nodal precession rate of flat circumbinary discs and the inferred disc mass is not large enough for self-gravity to be important.

The surface density profile is initially a power law distribution given by
 \begin{equation}
     \Sigma(R) = \Sigma_0 \bigg( \frac{r}{r_{\rm in}} \bigg)^{-p},
     \label{eq::sigma}
 \end{equation}
where $\Sigma_0$ is the density normalization and $p$ is the power law index. Note that the density normalization is set from the total disc mass above  (the simulated total mass is similar to the amount of mass from the three dust rings inferred from \cite{Bi2020}, assuming a gas-to-dust ratio of $100$). We use a locally isothermal disc with a disc thickness that is scaled with radius as
 \begin{equation}
    H = \frac{c_{\rm s}}{\Omega} \propto r^{3/2-q}, 
 \end{equation}
where $\Omega = \sqrt{GM/r^3}$ and $c_{\rm s}$ is the sound speed. We choose $q = 0.5$ to ensure that $H/r = \rm const$ over the radial extent of the disc. We consider two values of the disc aspect ratio $H/r = 0.05,\, 0.1$ at $r = r_{\rm in}$. We take the \cite{shakura1973} $\alpha$ to be either $0.01$, $0.05$, or $0.1$. We use the $\alpha$ prescription detailed in \cite{Lodato2010} given as
\begin{equation}
\alpha \approx \frac{\alpha_{\rm AV}}{10}\frac{\langle h \rangle}{H},
\end{equation}
where $\alpha_{\rm AV}$ is the artificial viscosity and $\langle h \rangle$ is the mean smoothing length on particles in a cylindrical ring at a given radius \citep{Price2018}.  The viscosity and mean smoothing length are not constant over the disc because we set the disc aspect to be constant.  Table~\ref{table::A1} shows the minimum and maximum values for the $\alpha$ viscosity parameter and the shelled-averaged smoothing length per scale height  $\langle h \rangle/H$.


\subsection{Adding a planet}
\label{sec::planet_setup}
We also consider two simulations with a  planet that is inclined to the binary orbit but coplanar to the initial disc. The disc has the same surface density profile from eq.~\ref{eq::sigma} but we implement a pre-carved gap in the disc in order to prevent excessive accretion of material on to the planet \citep[see for example][]{Lubow2016,Martin2016}. The inner and outer boundaries of the gap are taken to be $56\, \rm au$ and $153\, \rm au$, which are taken  from observations \cite[e.g.,][]{Bi2020,Kraus2020}. The initial semi-major axis of the planet is set to be roughly at the center of the gap between the inner and middle rings at $\sim 100\, \rm au$. We  consider a planet mass of $M_{\rm p} = 1\, \rm M_{J}$. This mass is sufficient enough to open a gap in the gas \citep{Lin1986,Marsh1992,Nelson2000}.  The dynamics are qualitatively the same regardless of planet mass.  We keep the viscosity constant across these two simulations ($\alpha \approx 0.01$), however, the gap opening is dependent on the viscosity meaning that at lower viscosity planets are able to open gaps easier \cite[e.g.,][]{Duffell2013}. The planet has an initially circular orbit with an accretion radius of $0.25\, \rm r_{\rm H} = 3.82\, \rm au$ \citep[e.g.][]{Nealon2018b}, where $r_{\rm H}$ is the Hill radius.  The pre-carved gap can be seen in the first and third panels in Fig.~\ref{fig::planet8_splash} for the two different disc aspect ratios, $0.1$ and $0.05$. The initial surface density profile of the pre-carved gap for $H/r = 0.05$ is given by the black line in the upper panel of Fig.~\ref{fig::disc_params_planet}.

\subsection{CO maps}

The hydrodynamical simulations do not include dust grains. Instead we simply assume that the dust distribution contains small, well coupled grains which represents the distribution of the gas. We construct a dust population with a grain-size distribution $dn/ds \propto s^{-m}$ between $ s_{\rm min} = 0.03$ and $ s_{\rm max} = 1000\, \mu m$ with $m = 3.5$. We assume the gas-to-dust ratio value of $100$ and calculate the total mass of dust from the gas mass in the simulation. The dust grain opacities are calculated assuming spherical and homogeneous grains (according to Mie theory) and are temperature independent, and we assume astrosilicates composition \citep{Weingartner2001}.  The stars are both assumed to be $1\, \rm Myr$ old and the mass of the stars comes directly from the simulation. The stellar luminosities are calculated using isochrones from \cite{Siess2000}. We assume that the dust and gas are in thermal equilibrium given that the circumbinary disc is passively heated. We use $10^8$ photon packets on a Voronoi mesh built directly on the particle distribution.  To calculate the moment maps, we additionally assume a uniform abundance ratio of $^{12}\rm CO$-to-$\rm H_2$ of $10^{-4}$ and we use $80\, \rm m/s$ resolution. Consistent with \cite{Bi2020}, our synthetic maps are convolved with the ALMA CLEAN beam of 0.122" $\times$ 0.159" (the beam is shown in the lower left of the synthetic maps).

\section{Results}
\label{sec::results}

Here we show the results of the hydrodynamical simulations.  First, we compare the disc evolution around a triple star system versus a binary system. Second, we simulate a disc around a binary (rather than the triple, see Section~\ref{sec::three-body}) and compare the two sets of system parameters from \citet{Czekala2017} and \citet{Kraus2020}.  Third, we examine in detail the evolution of the tilt, longitude of the ascending node and surface density profile in a disc without a planet.  Fourth, we introduce a giant planet at 100~au and again consider the disc evolution.  Finally we compare our results to the recent work by \citet{Kraus2020}.

When we analyse the SPH simulations, we separate the disc into 300 radial bins that span from the inner-most bound particle to the initial outer disc radius. Within each bin, we calculate the azimuthally averaged surface density, longitude of ascending node, tilt, twist, and eccentricity. 
 The tilt, $i$, is defined as the angle between the initial angular momentum vector of the binary (the $z$-axis) and the angular momentum vector of the disc.
 The twist, $\phi$, is measured  relative to the $x$-axis (the initial binary eccentricity vector).

\begin{figure}
 \centering
  \includegraphics[width=\columnwidth]{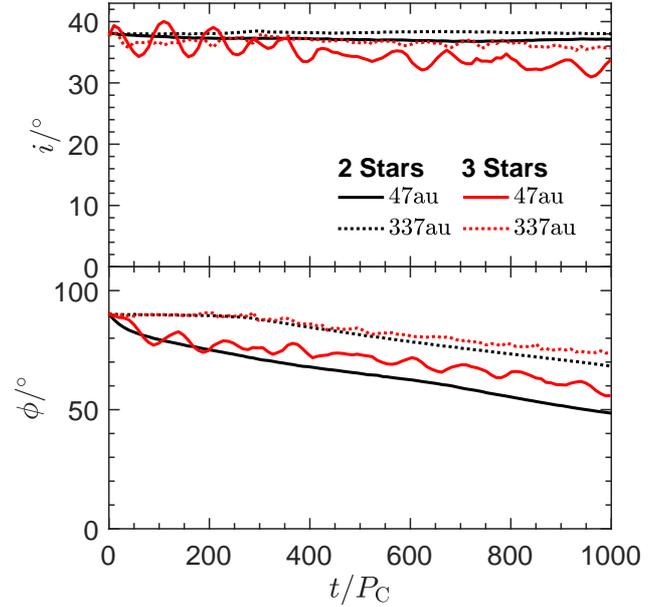}
  \caption{  Evolution of the inclination $i$, and longitude of the ascending node $\phi$, both as a function of time in units of $P_{\rm C}$, where $P_{\rm C}$ is the orbital period of star C. The disc viscosity is $\alpha = 0.01$. The black lines represent a circumbinary disc (run6 from Table~\ref{table::setup}), while the red lines represent a circumtriple disc (run0).  The disc is evaluated at two radii, $45\, \rm au$ (solid) and $180\, \rm au$ (dotted). The evolution of a circumbinary disc around GW Ori is similar to a circumtriple disc.} 
\label{fig::3star_compare}
\end{figure}

\begin{figure}
 \centering
  \includegraphics[width=\columnwidth]{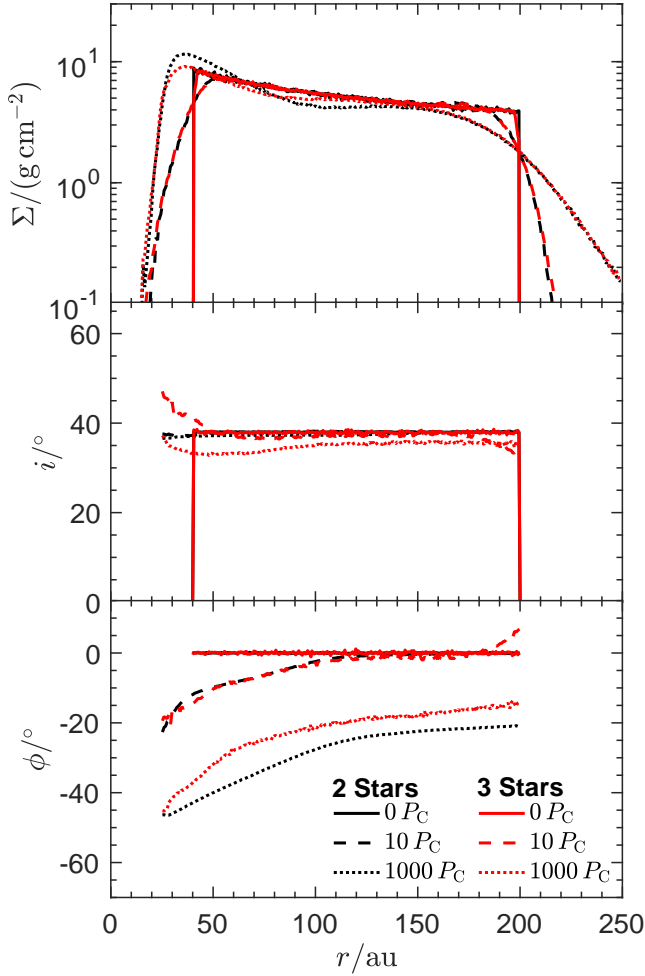}
  \caption{  Evolution of the disc surface density $\Sigma$ (top panel), tilt $i$ (middle panel), and twist $\phi$ (bottom panel), as a function of radius for a disc viscosity $\alpha = 0.01$. The black lines represent a circumbinary disc (run6 from Table~\ref{table::setup}), while the red lines represent a circumtriple disc (run0). We evaluate the disc at three different times, $t = 0\, P_{\rm C}$ (solid), $ 10\, P_{\rm C}$ (dashed), and  $1000\, P_{\rm C}$ (dotted), where $P_{\rm C}$ is the orbital period of star C. The tilt and twist are only shown in the range from $25\, {\rm au} < r < 200\, \rm au$. There is less mass in the inner parts of the circumtriple disc than the circumbinary disc and the warp profiles are similar for the two discs.}
\label{fig::3star_sigma}
\end{figure}

  \begin{figure}
 \centering
  \includegraphics[width=\columnwidth]{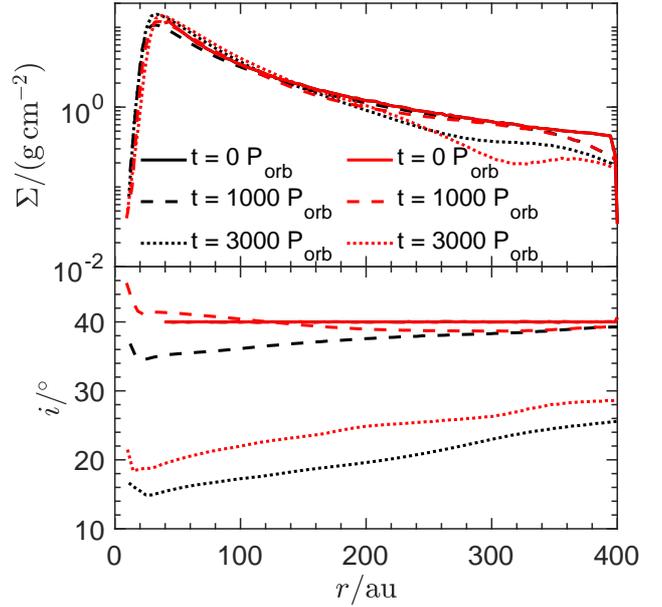}
  \caption{The  disc surface density profile ($\Sigma$, upper panel) and disc tilt ($i$, lower panel) as a function of disc radius for runs 1 and 2 from Table~\ref{table::setup}. The black lines correspond to a simulation with the binary parameters of \citet{Czekala2017} and the red lines represent a simulation with the binary parameters of \citet{Kraus2020}. The line style denotes the time at which the disc measurements are taken, with the solid, dashed, and dotted corresponding to times $t = 0$, $1000$, and $3000\, \rm P_{orb}$, respectively. The surface density profile and tilt evolution shows similar structures independent of the binary parameters used. }
\label{fig::disc_compare}
\end{figure}


 \subsection{Three-star hydrodynamical simulations}
 \label{sec:triple}
  In this section, we further compare the results of our $N$--body calculations of modeling a triple star system as a binary star system by using hydrodynamical simulations. We use the stellar and disc parameters from \cite{Kraus2020} from Table~\ref{table::binaryparams}, however, we model our circumtriple disc at a higher-resolution ($10^6$ Lagrangian particles). Figure~\ref{fig::3star_compare} shows the evolution of the disc tilt and longitude of the ascending node as a function of time for a disc viscosity $\alpha = 0.01$ (bending-wave regime). The disc tilt follows a similar evolution regardless of the stellar prescription. The outer parts of the discs (dotted lines) are closer to each other in tilt than inner parts of the discs (solid lines). However, the tilt changes are only a few degrees.
 
 
We do not see any evidence for the disc breaking due to the tidal torque of the triple star, or the binary star. The precession of the circumtriple disc is similar to that of the circumbinary disc 
(lower panel in Fig.~\ref{fig::3star_compare}). 
Figure~\ref{fig::3star_sigma} compares the surface density, tilt, and twist profiles for the circumbinary and circumtriple discs at three different times. The surface density profiles at all times for both discs are
smooth, confirming that neither disc is broken.  
The warp in the circumtriple disc is consistent with the warp profile in the circumbinary disc. Lastly, the precession rate of the circumtriple disc as a function of radius is slightly slower at later times than the circumbinary disc. 
This suggests that the circumtriple disc is more stable against breaking than the circumbinary disc for the parameters of these simulations.

With the above comparison, we are confident that we can model the GW Ori circumtriple disc as a circumbinary disc. With this assumption, we can explore a larger span of the parameter space while increasing computational efficiency. The remaining simulations discussed in this work will be modelling a circumbinary disc.

\subsection{Binary parameters from \citet{Czekala2017} vs. \citet{Kraus2020}}
Recently, \cite{Kraus2020} presented additional high resolution observations of the  GW Ori circumtriple system.  Before their updated stellar parameters became public, we used the stellar parameters presented in \cite{Czekala2017} to approximate GW Ori as a binary system. Here, we test the difference in the disc evolution between using the binary parameters from \cite{Czekala2017} compared to \cite{Kraus2020}. 

Figure~\ref{fig::disc_compare} shows the  disc surface density profile (upper panels) and disc tilt (lower panel) as a function of disc radius for runs 1 and 2 from Table~\ref{table::setup}. The disc surface density profiles are similar irrespective of the binary parameters used. There is $\lesssim 20\%$ difference in the tilt evolution and the shape of the tilt profile is similar in both models.
Based on this, we conclude that the disc will evolve in a similar fashion when using either set of binary parameters. Thus, unless otherwise indicated, we use the binary parameters given by \citet{Czekala2017}.


\begin{figure*}
 \centering
  \includegraphics[width=0.9\columnwidth]{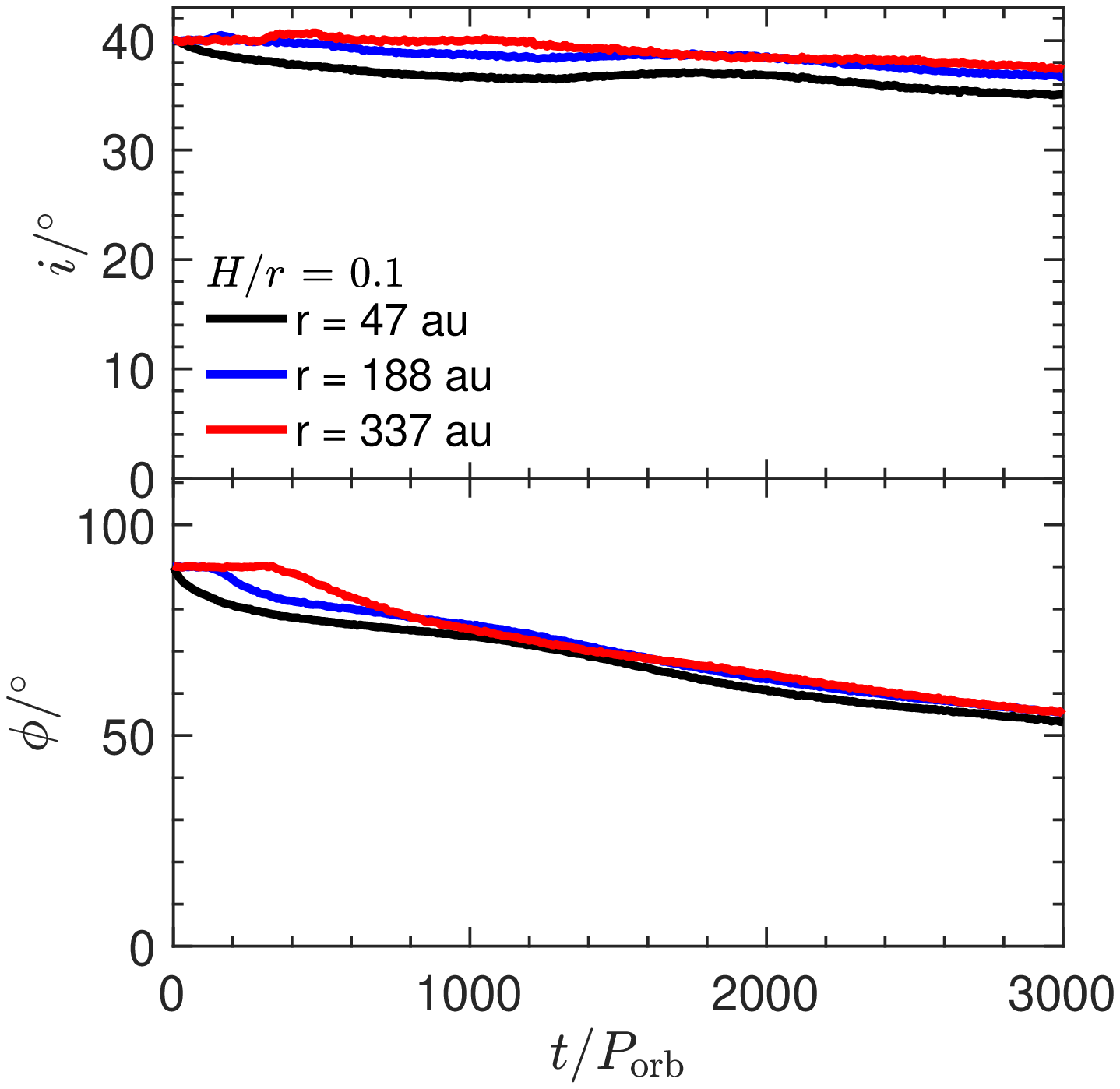}
  \includegraphics[width=0.9\columnwidth]{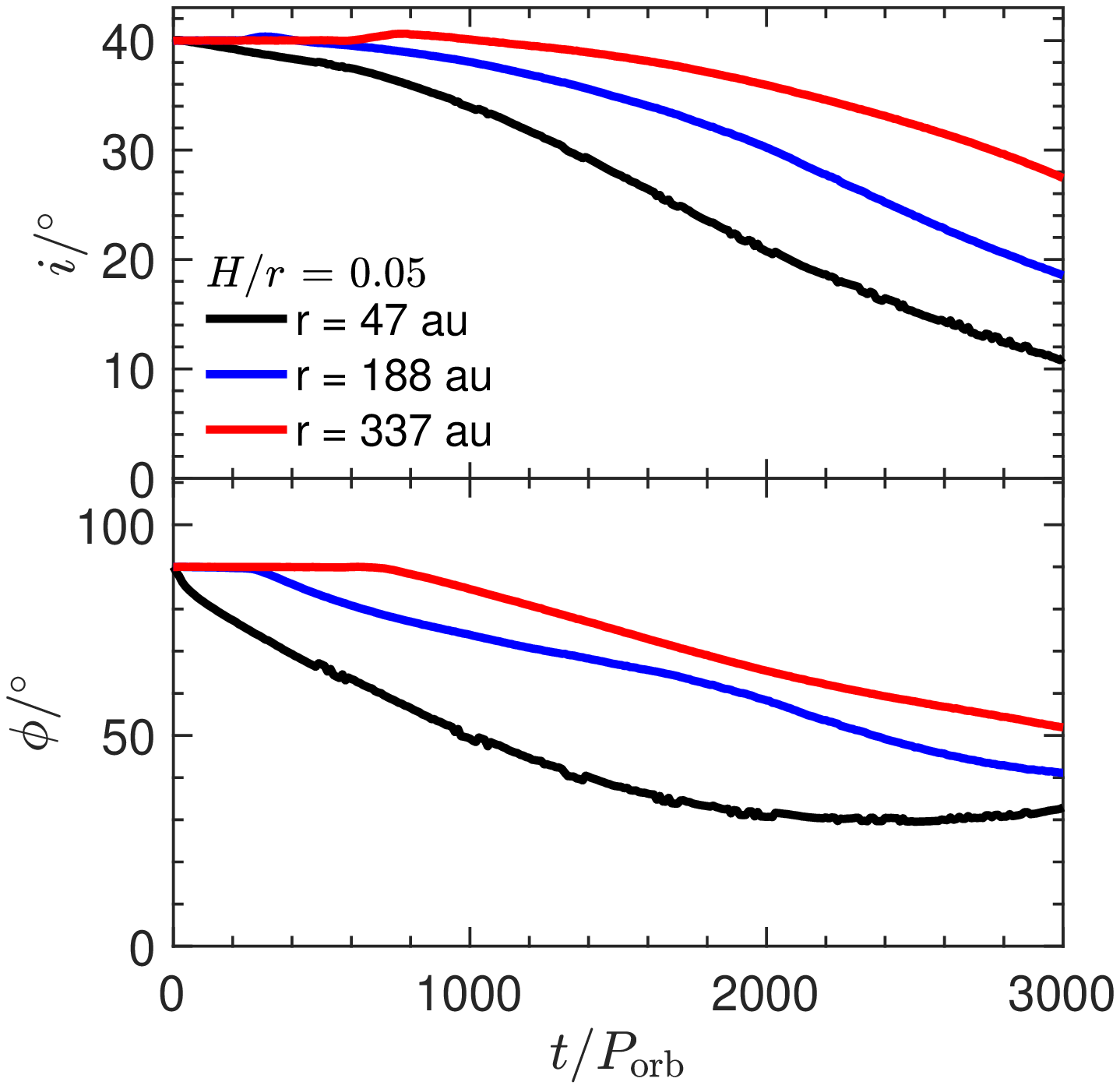}
  \caption{ Evolution of the inclination, $i$, and longitude of the ascending node, $\phi$, both as a function of time for two different disc aspect ratios. Left panel: $H/r = 0.1$ (run3 from Table~\ref{table::setup}). Right panel: $H/r = 0.05$ (run1). The disc is  evaluated at three radii, $47\, \rm au$ (black), $188\, \rm au$ (blue), and $337\, \rm au$ (red). A thinner disc shows strong warps when compared to a thicker disc.}
\label{fig::time_plots_disc}
\end{figure*}

\begin{figure*}
  \includegraphics[width=1.8\columnwidth]{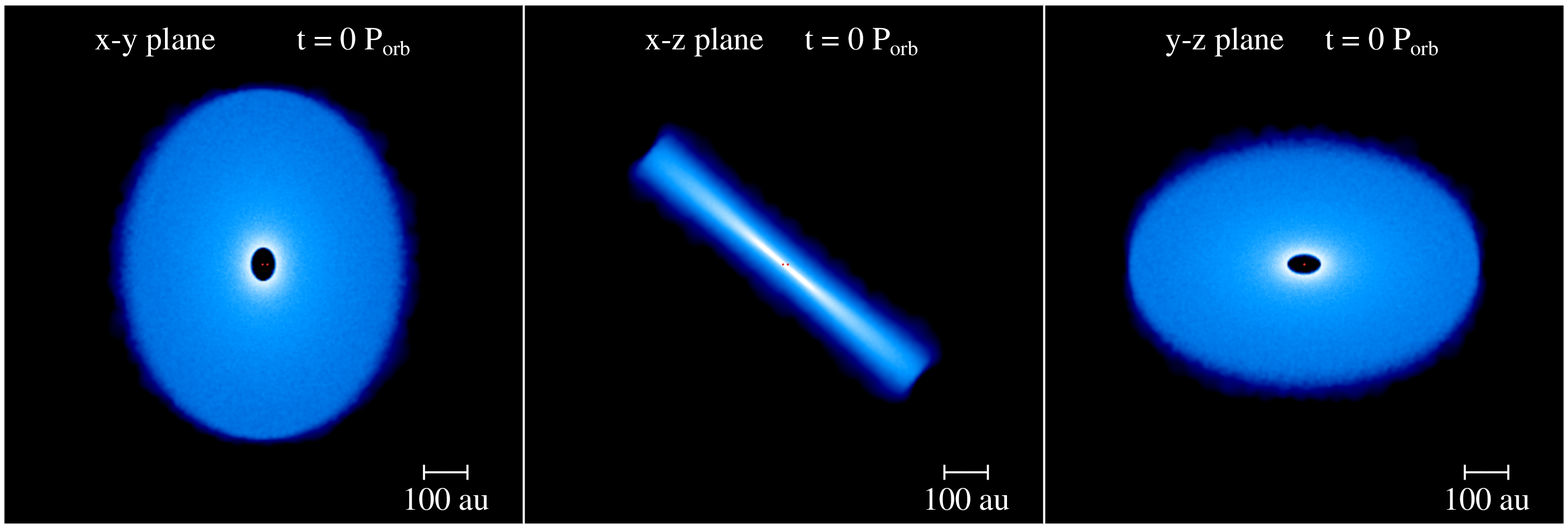}
  \includegraphics[width=1.8\columnwidth]{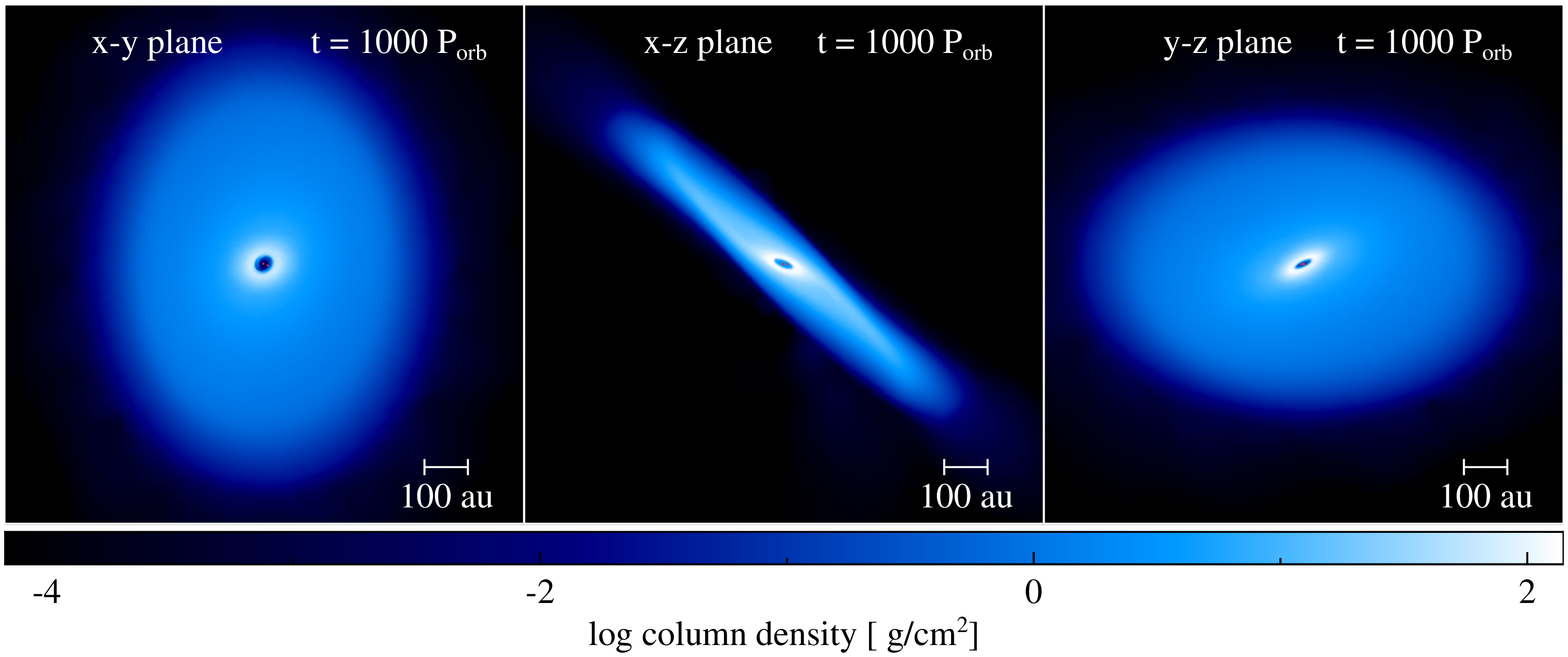}\centering
  \caption{Disc evolution for a circumbinary disc with $i_0 = 40\degree$ with a disc aspect ratio $H/r = 0.05$ (run1 from Table~\ref{table::setup}). Upper panels: initial set-up for the GW Ori disc around an eccentric binary with separation of $9.2\, \rm au$. The bottom panels: the disc at a time of $t = 1000\, \rm P_{orb}$.  The colour bar denotes the gas density. The left-hand panels show the view looking down on to the binary orbital plane, the $x$--$y$ plane. The middle panels shows the $x$--$z$ plane and the right-hand panels show the $y$--$z$ plane. The binary torque causes the disc to become strongly warped but does not break. }
\label{fig::disc_splash}
\end{figure*}

\begin{figure}
  \includegraphics[width=\columnwidth]{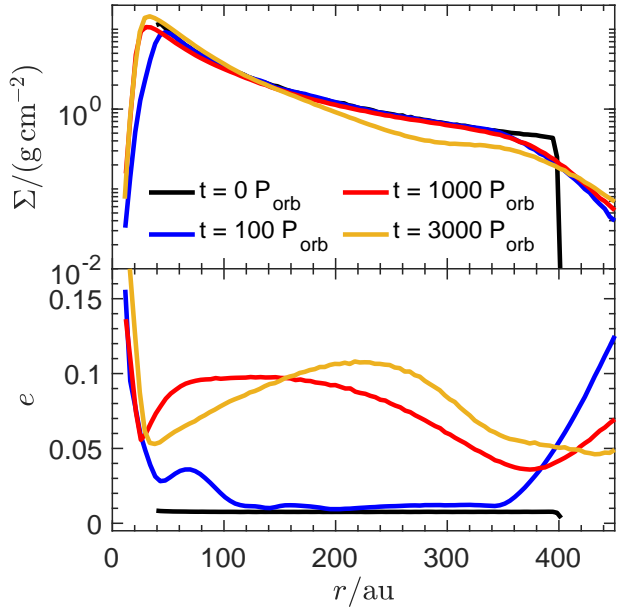}
  \caption{Surface density $\Sigma$ (top panel) and eccentricity $e$ (bottom panel) as a function of radius at different times for the $H/r = 0.05$ simulation (run1 from Table~\ref{table::setup}) shown in Figure 4. The black, blue, red, and yellow curves  correspond to $t = 0, 100, 1000, 3000\, \rm P_{orb}$, respectively. The disc maintains a smooth surface density profile which is indicative of no disc breaking.}
\label{fig::disc_params}
\end{figure}

\begin{figure*}
  \includegraphics[width=\columnwidth]{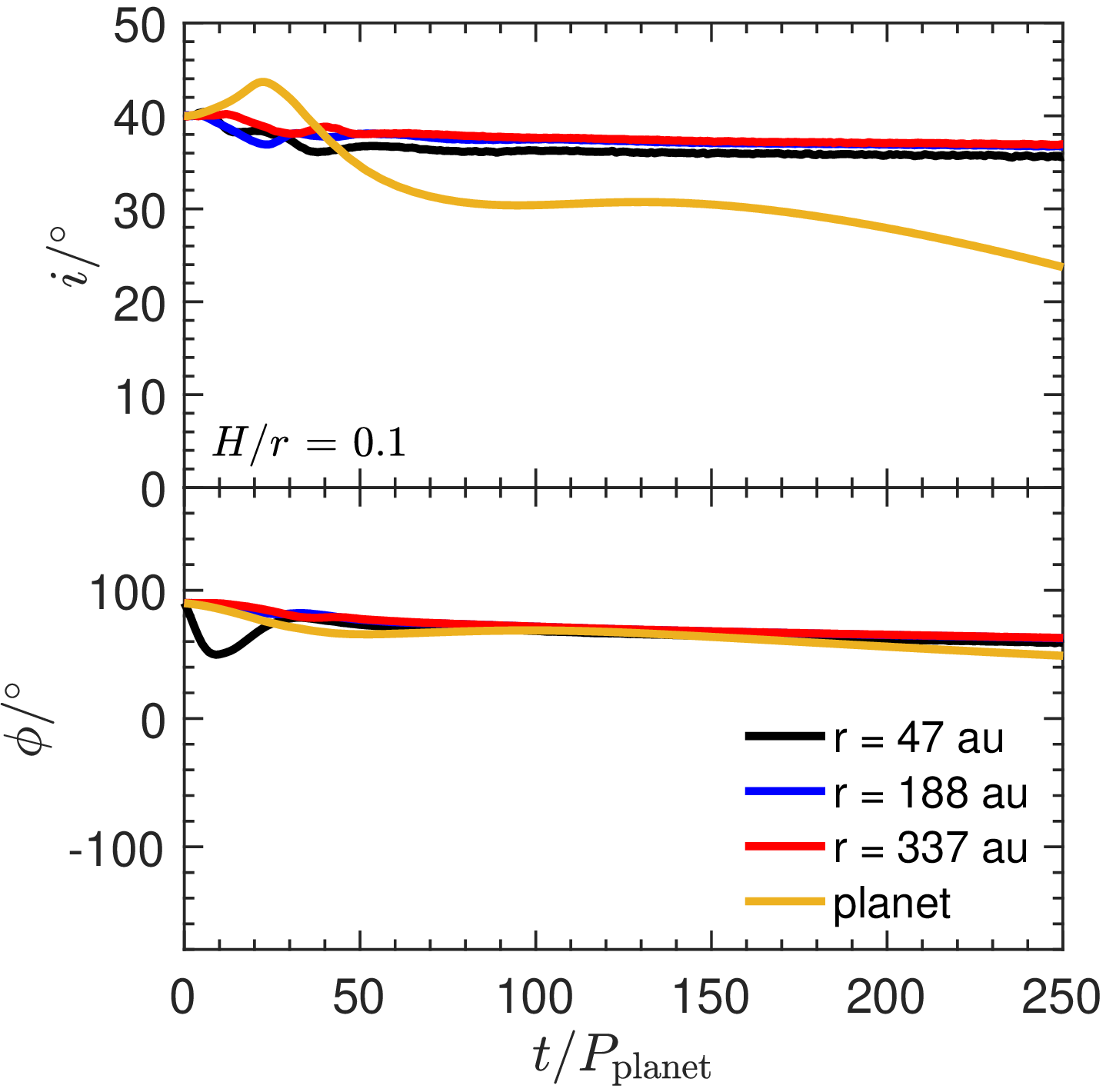}
  \includegraphics[width=\columnwidth]{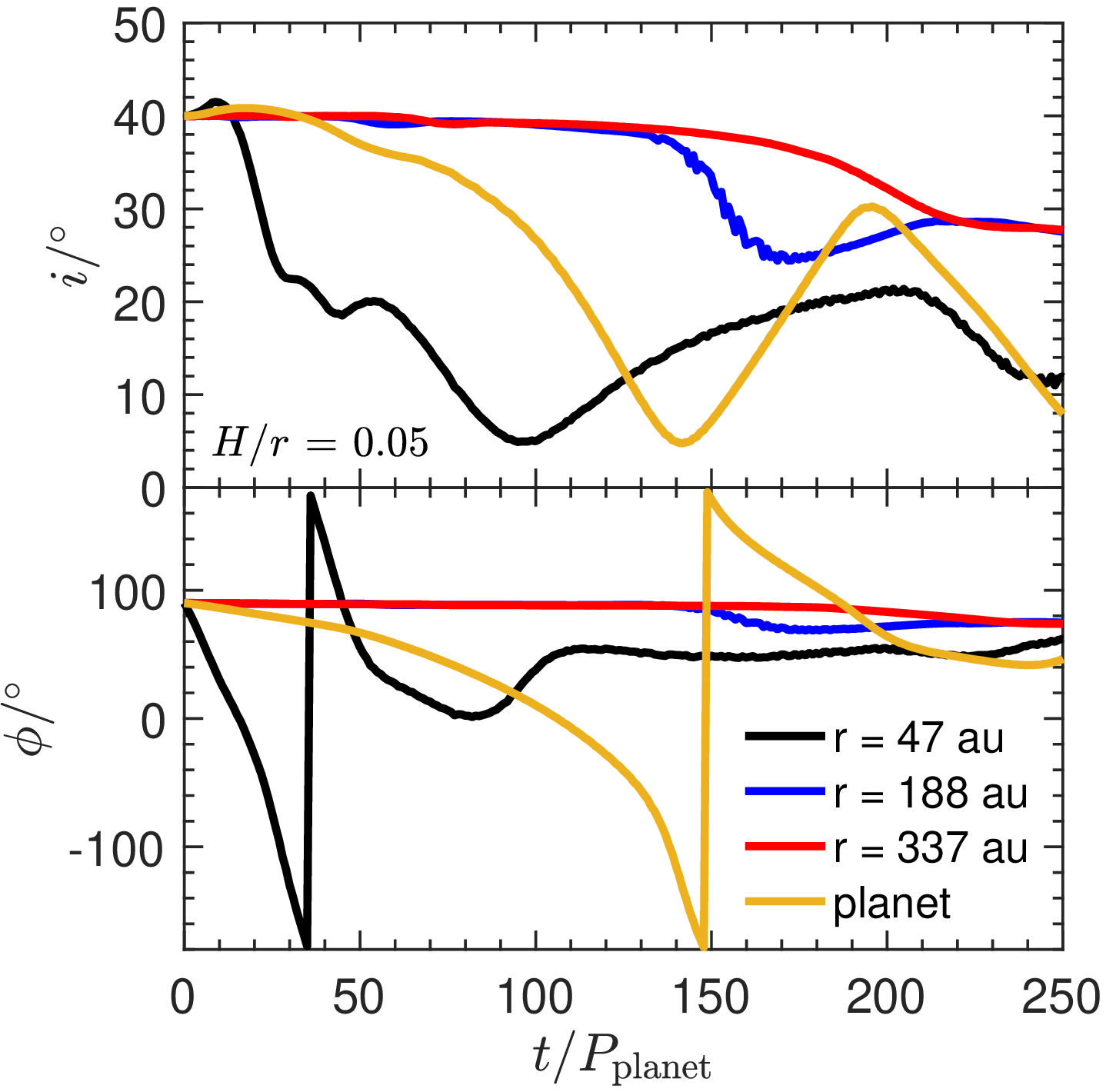}\centering
  \caption{ Evolution of the inclination, $i$, and longitude of the ascending node, $\phi$, both as a function of time for two different disc aspect ratios. Left panel: $H/r = 0.1$ (run5 from Table~\ref{table::setup}). Right panel: $H/r = 0.05$ (run4). The disc is  evaluated at three radii, $47\, \rm au$ (black), $188\, \rm au$ (blue), and $337\, \rm au$ (red). The planet is given by the yellow lines. 
  }
\label{fig::time_plots_planet}
\end{figure*}

\begin{figure*}
  \includegraphics[width=0.5\columnwidth]{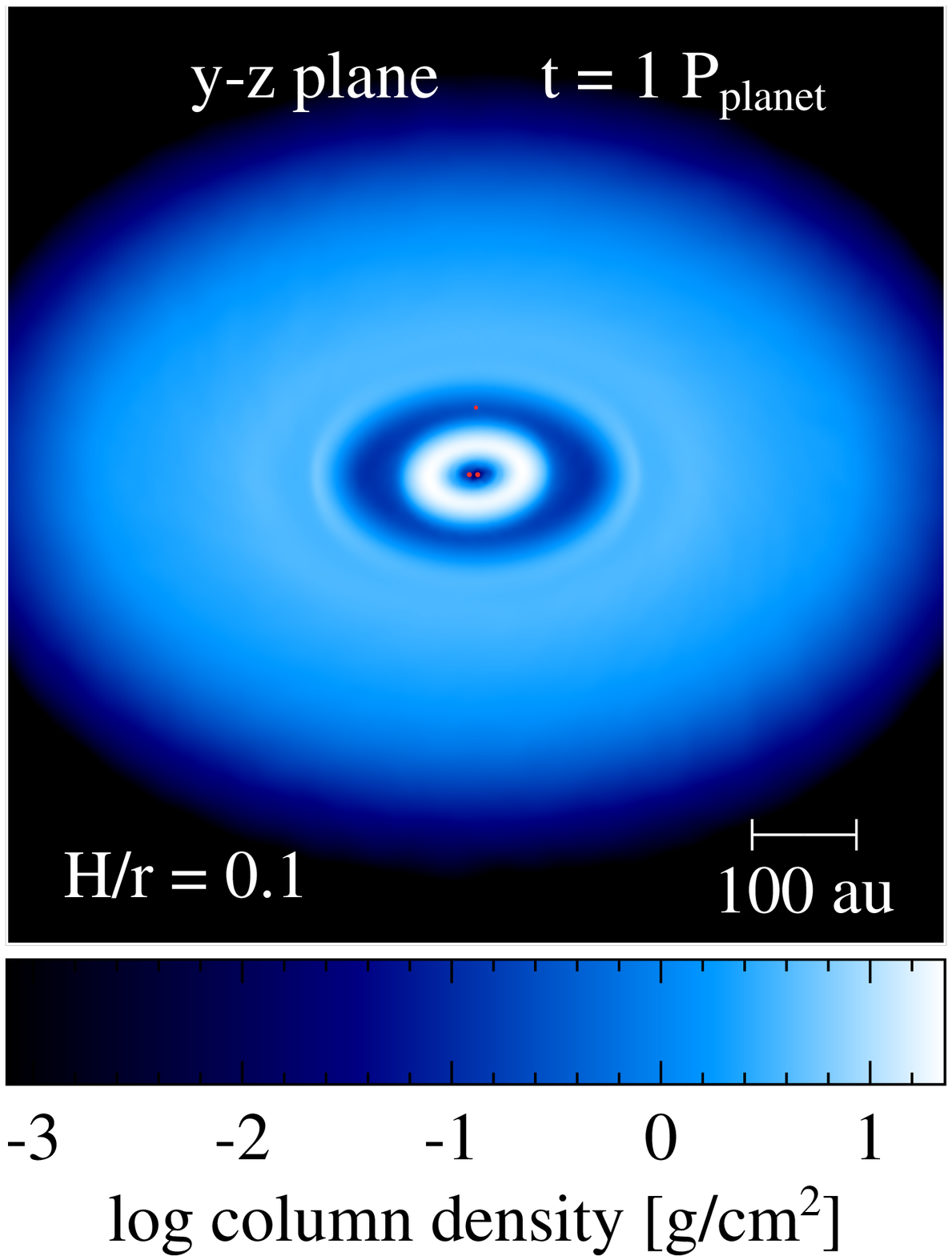}
  \includegraphics[width=0.5\columnwidth]{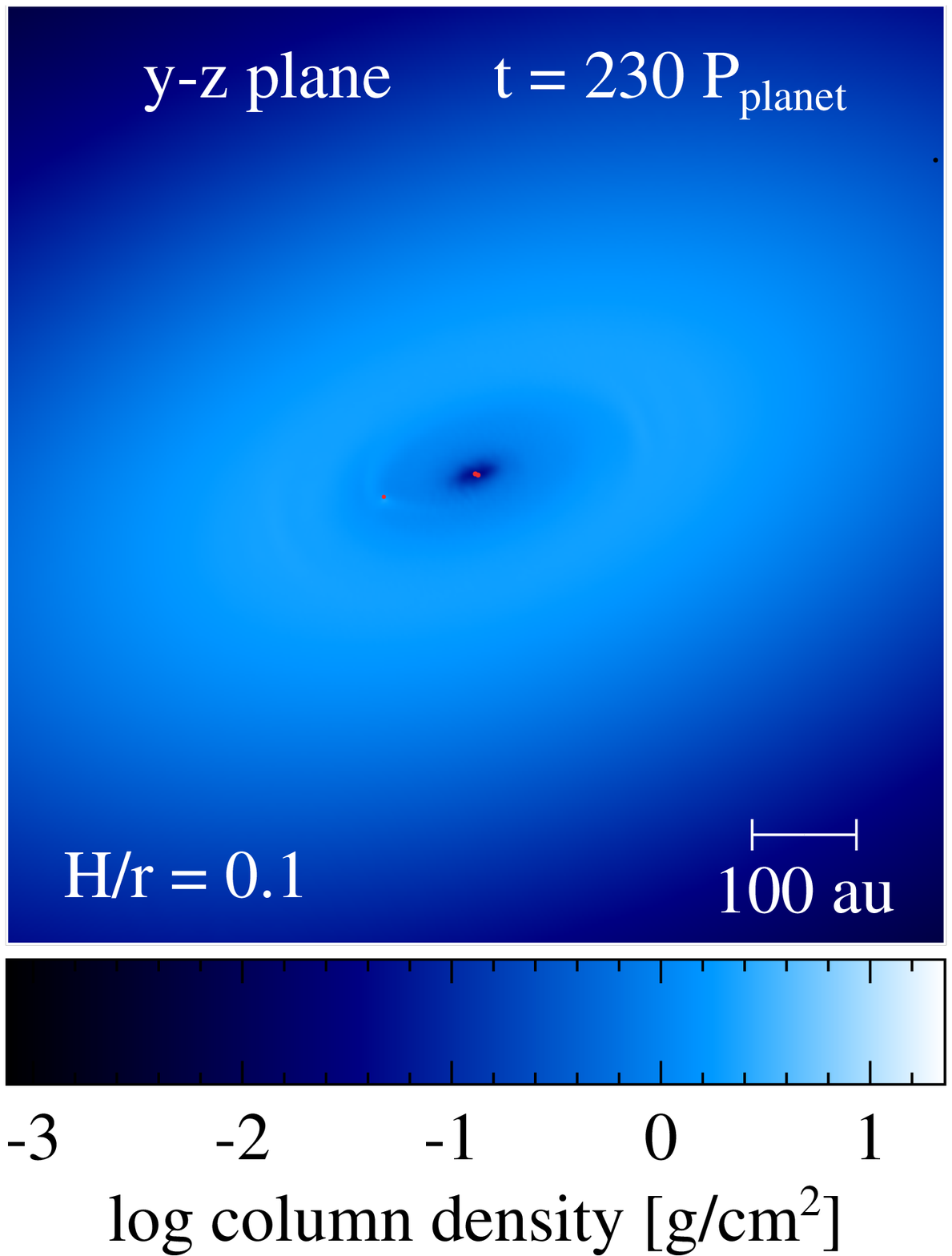}
  \includegraphics[width=0.5\columnwidth]{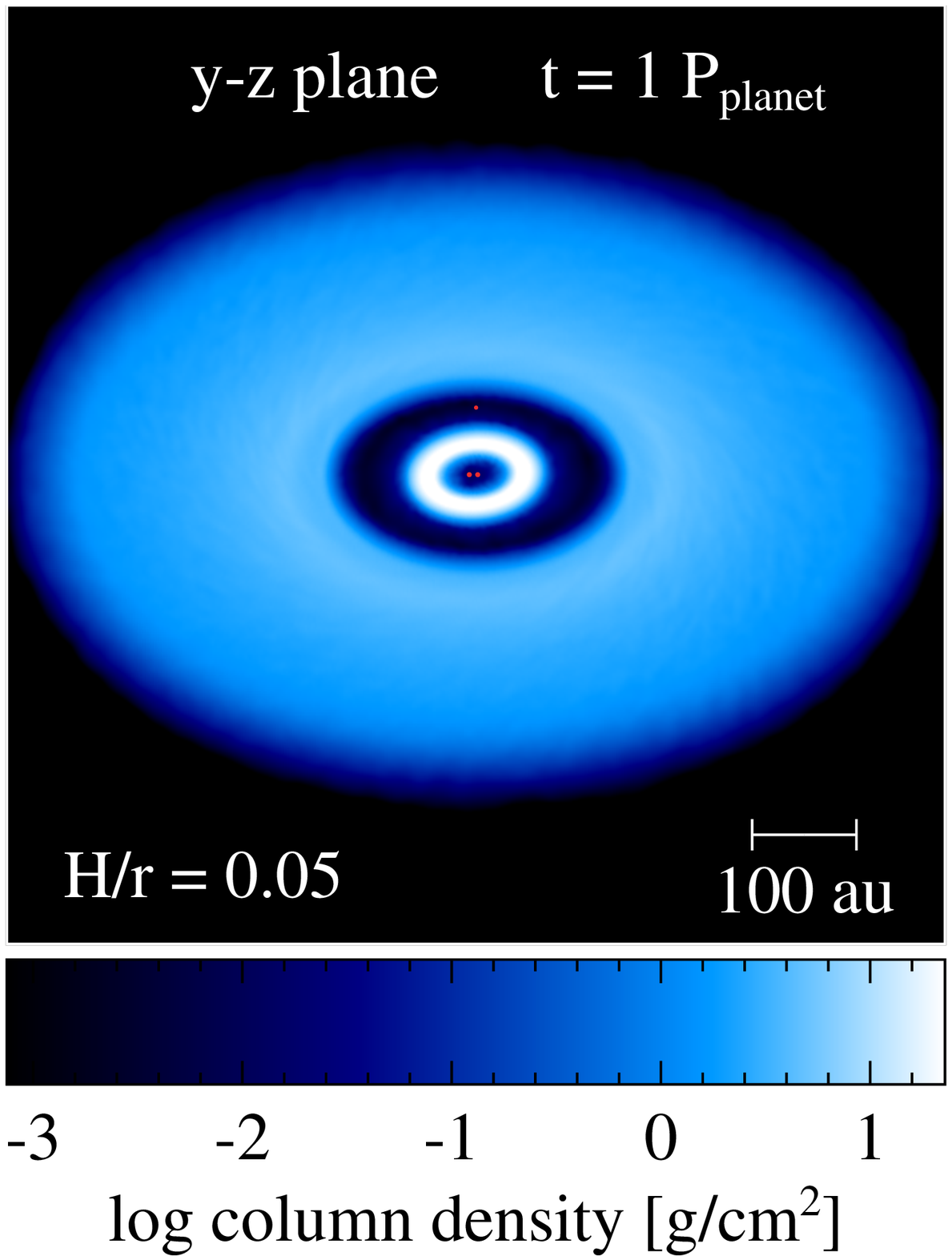} 
  \includegraphics[width=0.5\columnwidth]{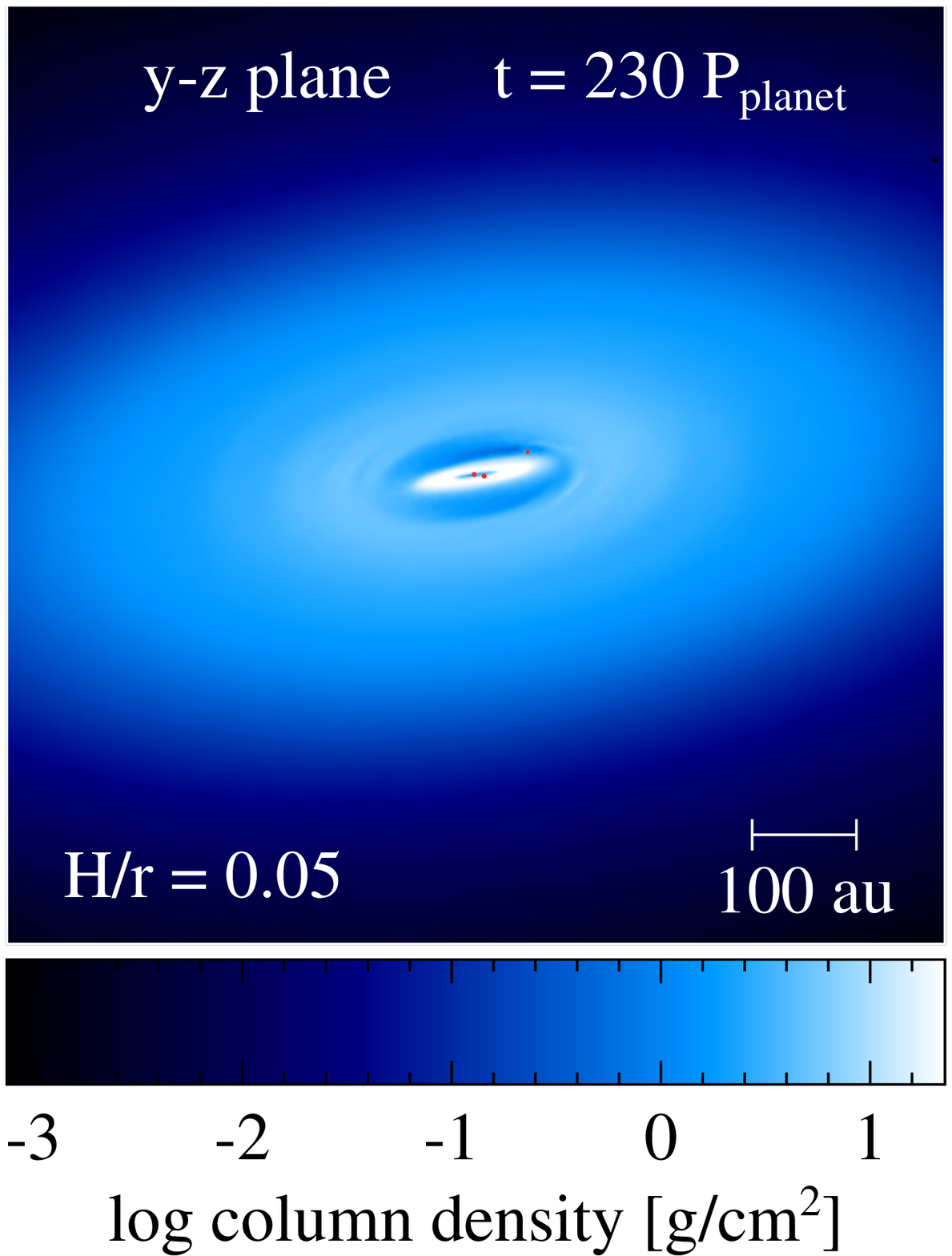}\centering
  \caption{Disc evolution for a circumbinary disc with $i_0 = 40\degree$ along with a circumbinary planet with $i_{0,\rm p} = 40\degree$. We show the results for two different aspect ratios $H/r = 0.1$ (two left-most panels, run5 from Table~\ref{table::setup} ) and $0.05$ (two right-most panels, run4). The first and third panel beginning from the left shows the evolution at a time of $t = 1\, \rm P_{planet}$ with the pre-carved gap for $H/r = 0.1$ and  $H/r = 0.05$. The second and fourth panels from the left, shows the disc evolution at a time of $t = 230\, \rm P_{planet}$ for the two different disc aspect ratios.The colour bar denotes the gas density. We show the view in the $y$--$z$ plane. A planet is able to carve a gap within a thin disc. }
\label{fig::planet8_splash}
\end{figure*}


\subsection{Effect of disc parameters}
\label{sec::woutaplanet}
Here we test the dynamical effects the binary has on two different disc aspect ratios, $H/r = 0.1$ and $0.05$. The upper of these values represents the thicker aspect ratio that has been inferred for GW Ori \citep{Bi2020} while the lower is a more typical value expected for protoplanetary discs \citep[e.g.][]{DAlessio1998}. 

The left panel of  Fig.~\ref{fig::time_plots_disc} shows the evolution of the disc inclination and longitude of the ascending node for a disc aspect ratio $H/r = 0.1$ (run3 from Table~\ref{table::setup}). We probe the disc at three different radii, $47$, $188$, and $337\, \rm au$ corresponding to the centres of the three observed dust rings (Table~\ref{table::rings}). 
 The tilts of the rings  show little warping and a slow alignment towards the binary orbital plane during the simulation.
From the evolution of the longitude of the ascending node, the disc shows a slow precession rate. Furthermore, the three measured radii are precessing at  roughly the same rate, which is further evidence that no disc breaking has occurred.

The right panel of  Fig.~\ref{fig::time_plots_disc} shows the evolution of the GW Ori disc with a smaller disc aspect ratio $H/r = 0.05$ (run1 from Table~\ref{table::setup}). Unlike the thicker disc, the thinner disc is more prone to warping (and breaking since the disc communication timescale is longer). The disc tilt (upper, right panel) decreases substantially across the entire disc as time increases. The lower sub-panel shows the longitude of the ascending node as a function of time. The three measured radii are precessing in a similar fashion which suggest the disc is not broken but strongly warped.  The decrease in the disc tilt in time is caused by the disc aligning to the orbital plane of the binary. Depending on the disc misalignment and binary eccentricity, due to dissipation the disc will evolve to one of two possible alignments. For sufficiently small initial inclination the disc precesses about the binary angular momentum vector and moves towards a coplanar alignment with the binary orbital plane \citep{papaloizou1995, Lubow2000,Nixon2011,Facchini2013,Foucart2014,Smallwood2019a}. The alignment timescale is very sensitive to the disc aspect ratio, $H/r$. For small $H/r$, the alignment timescale is shorter versus being longer for a more thicker disc \cite[e.g.,][]{Lubow2018,Smallwood2020}. Therefore, this is why the thicker disc in the left panel of Fig.~\ref{fig::time_plots_disc} is not aligning within our time domain, while the thinner disc in the right panel is aligning to the binary orbital plane.

 Figure~\ref{fig::disc_splash} shows the disc evolution for our simulation with the thinner aspect ratio of $H/r = 0.05$ (run1 from Table~\ref{table::setup}). The upper panels show the initial conditions where the disc is tilted by $40\degree$. The lower panels show the disc evolution at a time  $t = 1000\, \rm P_{\rm orb}$. The warp in the disc can be clearly seen in the middle panel and the continuous nature of the disc indicates that there is no break present in the disc.

To investigate the warping in further detail, in Fig.~\ref{fig::disc_params}, we show the surface density (top panel), and eccentricity (bottom panel) as function of disc radius.  As the disc evolves over time, the inner regions  viscously spreads inwards and the outer portions  outwards. The surface density profile at all times is smooth confirming that the disc is not broken. The bottom panel shows the disc eccentricity as a function of disc radius. The disc initially starts circular, however there is eccentricity growth that occurs as the warp propagates outwards.

We have demonstrated that the observed parameters in GW Ori do not lead to disc breaking. This motivates us to consider an alternative mechanism to provide a break in the disc, separating the disc into the distinct misaligned planes that are observed.

\begin{figure}
  \includegraphics[width=\columnwidth]{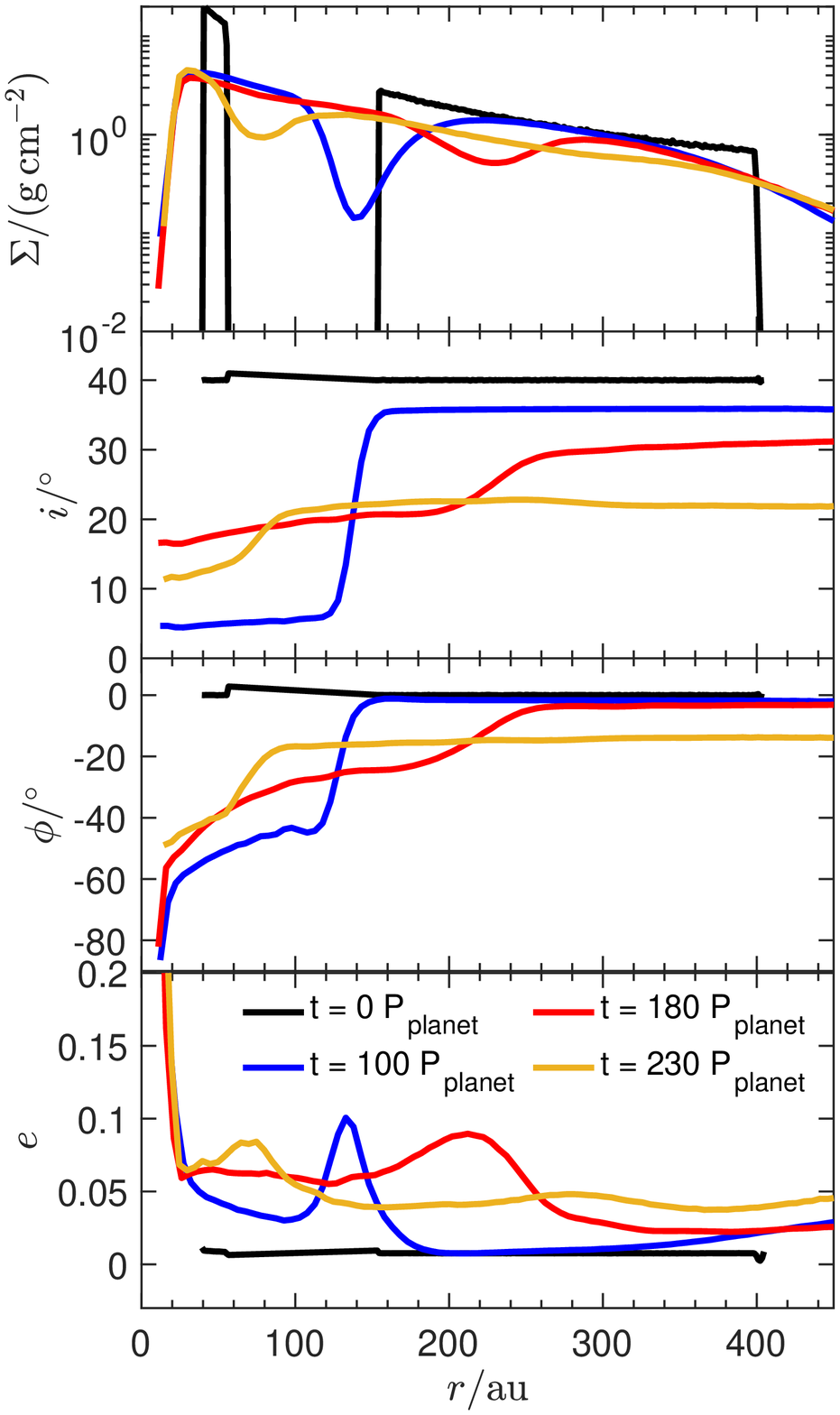}
  \caption{Beginning from the top we show the surface density $\Sigma$, disc tilt $i$, longitude of the ascending node $\phi$, and eccentricity $e$ as a function of radius at different times. The black, blue, red, and yellow curves  correspond to $t = 0, 100, 180, 230\, \rm P_{planet}$, respectively. The initial conditions for the circumbinary disc are for run4 from Table~\ref{table::setup}, which has a disc aspect ratio $H/r = 0.05$ and pre-carved gap. A planet is able to maintain a gap within the disc seen by the dips in the surface density profile. }
\label{fig::disc_params_planet}
\end{figure}

\begin{figure}
  \includegraphics[width=\columnwidth]{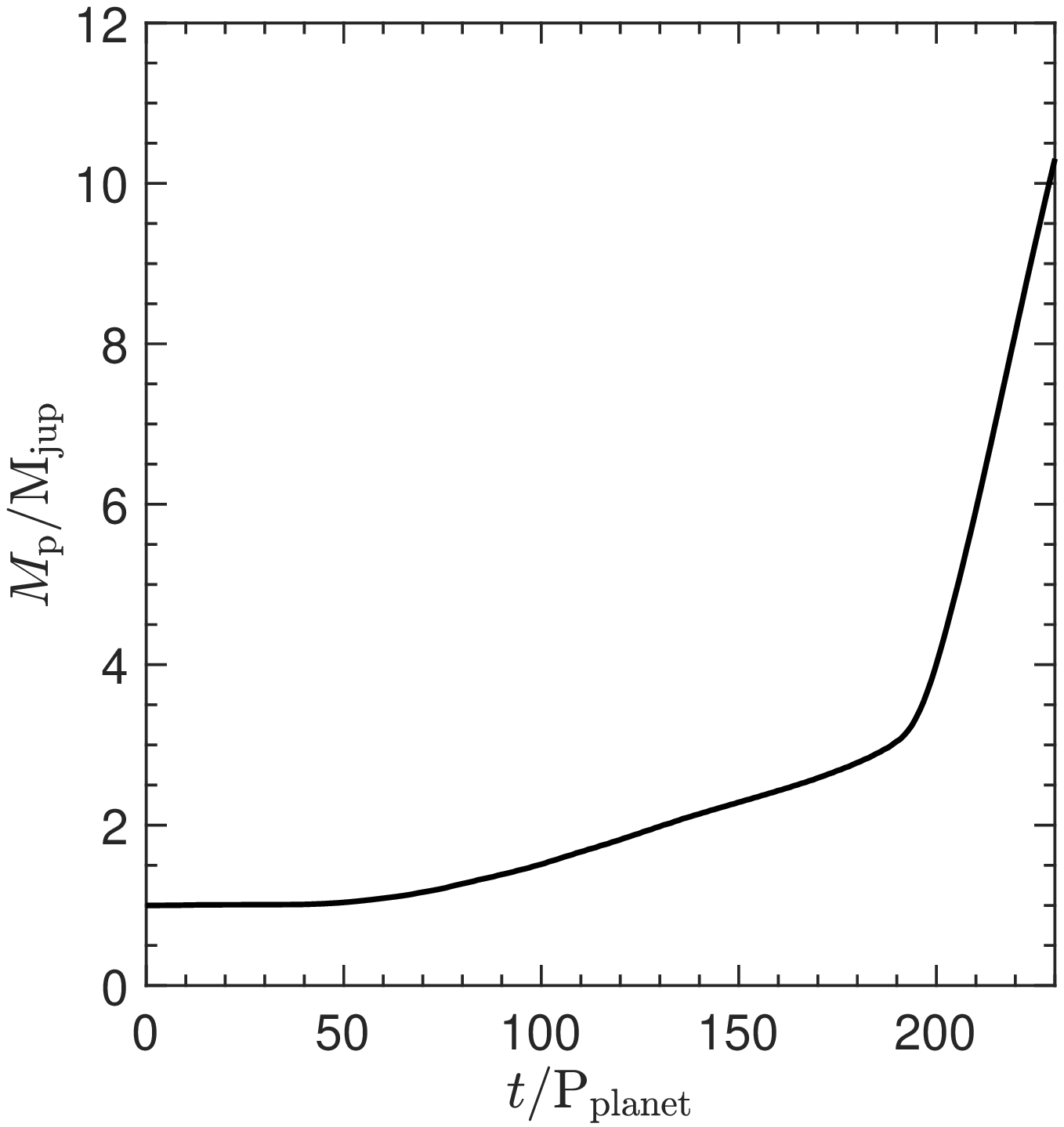}
  \caption{The planet mass, $M_{\rm p}$, as a function of time in planet orbital periods, $P_{\rm planet}$ from run5 from Table~\ref{table::setup}. The planet mass increases drastically after $t > 200\, \rm P_{planet}$.  }
\label{fig::planet_mass}
\end{figure}

\subsection{Effect of a giant planet}
\label{sec::planet}
In this section, we carry out SPH simulations with a planet that is initially coplanar with respect to the disc. Both the disc and the planet begin misaligned to the binary orbital plane. We again consider two disc aspect ratios, $H/r = 0.1$ (run5 from Table~\ref{table::setup}) and $0.05$ (run4), where $\alpha = 0.01$ in both simulations (such that both discs are in the wave-like regime). Due to the pre-carved gap (described in Section~\ref{sec::planet_setup}), initially the inner disc has a mass of $0.0133\, \rm M_{\odot}$ and the outer disc has $0.0867\, \rm M_{\odot}$. 
 
Figure~\ref{fig::time_plots_planet} shows the evolution of the tilt and longitude of the ascending node for disc aspect ratios $H/r = 0.1$ (left panel) and $H/r = 0.05$ (right panel).  We also include the evolution of the tilt and longitude of the ascending node for the planet. For the thicker disc simulation (left panel), the whole disc remains at an inclination similar to the initial conditions.  There is a small deviation in the twist of the disc in the inner parts because the simulation begins with a pre-carved gap which allows the inner disc to freely precess initially.  The initial planet evolution is dominated by the binary as it begins to undergo a tilt oscillation due to the binary eccentricity \citep[e.g.,][]{Smallwood2019a}. Once the planet tilt evolves out of the plane of the disc, the torque from the planet is no longer strong enough  to maintain the gap. The large disc aspect ratio means that viscous spreading of the disc occurs on a fast time-scale. Therefore, the gap then quickly fills with gas and the disc evolves rigidly.  The disappearance of the gap can be clearly seen in Fig.~\ref{fig::planet8_splash}, where the two left-most panels show the initial disc setup (with $H/r = 0.1$ and the pre-carved gap centered on $100\, \rm au$) and the disc evolution at a time $t = 230\, \rm P_{planet}$. The gap has  dissipated due to the faster disc spreading coupled with the misalignment of the planet to the disc. The later evolution of the planet is dominated by the interaction of the planet with the disc. Secular planet-disc tilt oscillations occur in discs in binaries \citep{Lubow2016,Martin2016}. In circumbinary discs around circular binaries, the planet tends to be closer to the binary orbital plane as a result of planet-disc interactions \citep{Pierens2018}.

The right panel of Figure~\ref{fig::time_plots_planet} shows the evolution for a thinner disc with $H/r = 0.05$. The lower disc aspect ratio means it is easier for the planet to create and hold open  a gap. Similar to the thick disc case, the planet becomes misaligned to the plane of the disc.  The evolution of the inner ring, centered at  $47\, \rm au$ (black lines), is dominated by tilt oscillations that are primarily driven by the outer disc. 
 The planet and the inner disc ring  undergo tilt oscillations that are driven by the outer disc. Both the inner ring and planet undergo similar evolution but on different timescales. The evolution of both are dominated by the massive outer disc part since their inclinations are lower than the outer disc \citep[e.g.][]{Pierens2018}. The middle ($188\, \rm au$) and outer ($337\, \rm au$) portions of the disc show evolutionary changes (decrease in inclination),  due to binary-disc alignment. Had we  not initially truncated the outer radius we would expect their inclinations to remain more constant.   Since this simulation has a lower disc aspect ratio than the simulation described in the previous paragraph, the viscous spreading of the disc in to the planet gap is slower which allows the planet to break the disc. The two right-most panels in Fig.~\ref{fig::planet8_splash} show the initial disc setup (with $H/r = 0.05$ and the pre-carved gap centered at $100\, \rm au$) and the disc evolution at a time $t = 230\, \rm P_{planet}$. The formation of a $1\, \rm M_{\rm J}$ planet in the disc is able to maintain a long lived strongly warped disc structure  when $h/r < 0.05$.

 \begin{figure*}
 \centering
  \includegraphics[width=1.1\columnwidth]{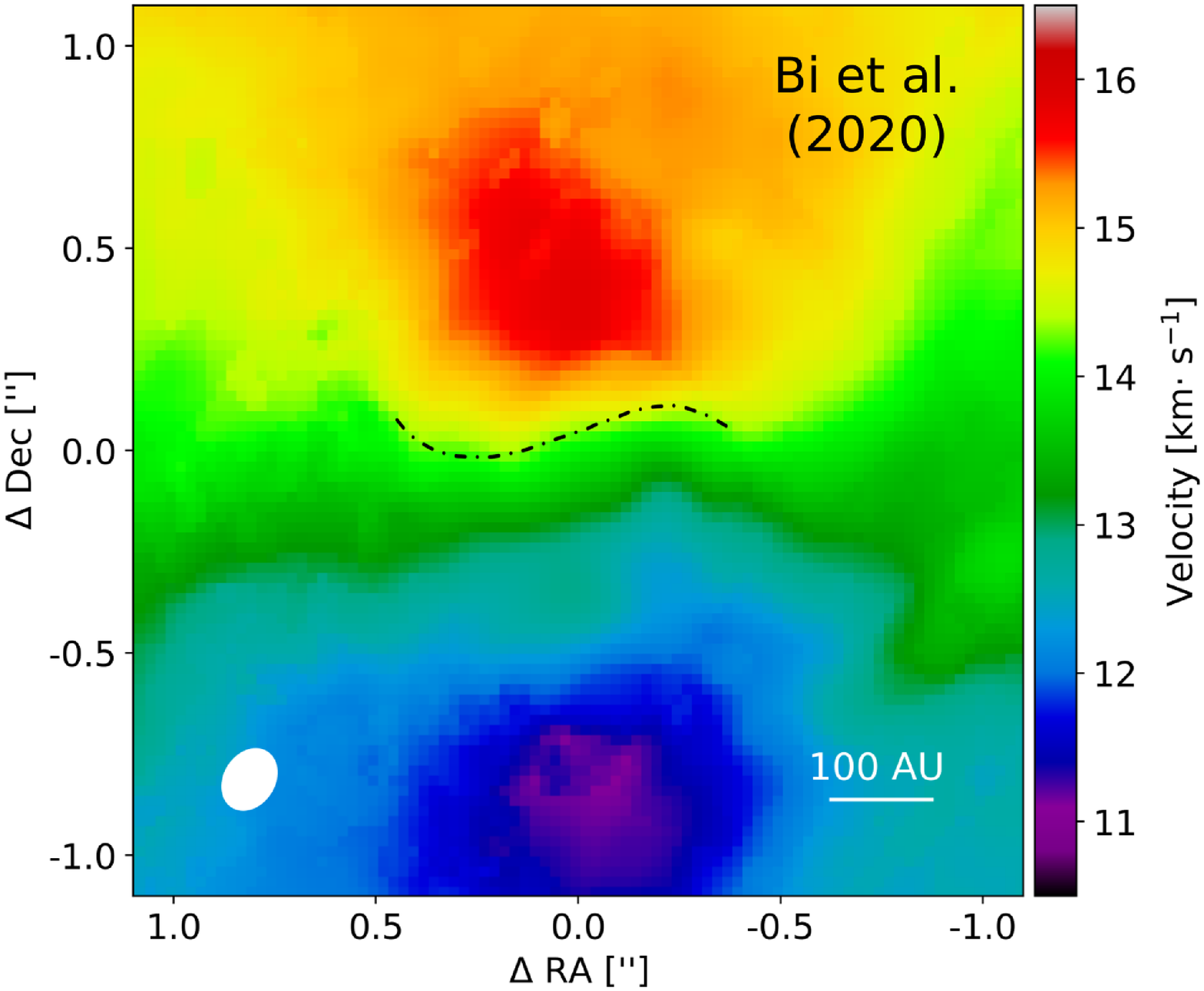}
  \includegraphics[width=\columnwidth]{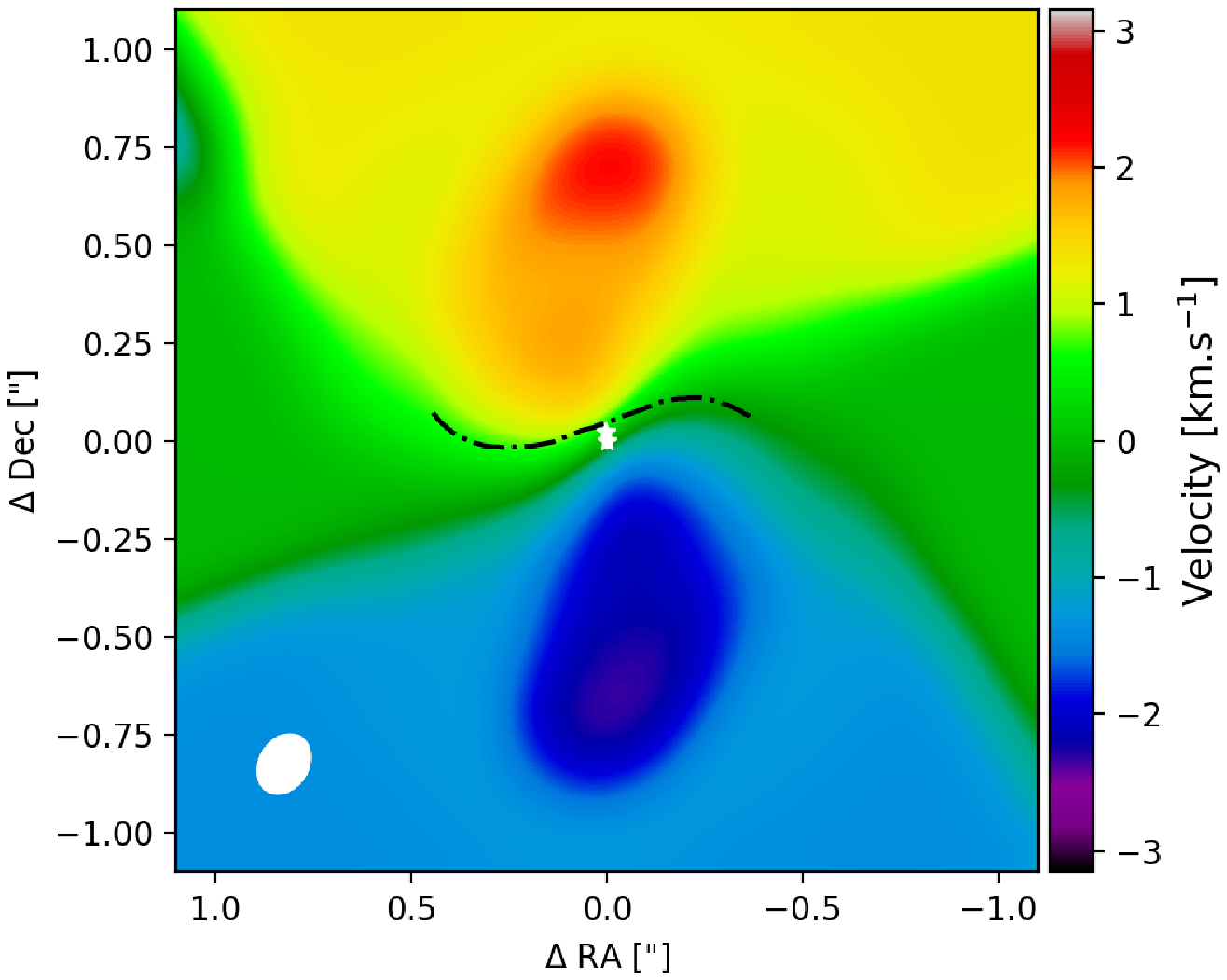}
  \includegraphics[width=\columnwidth]{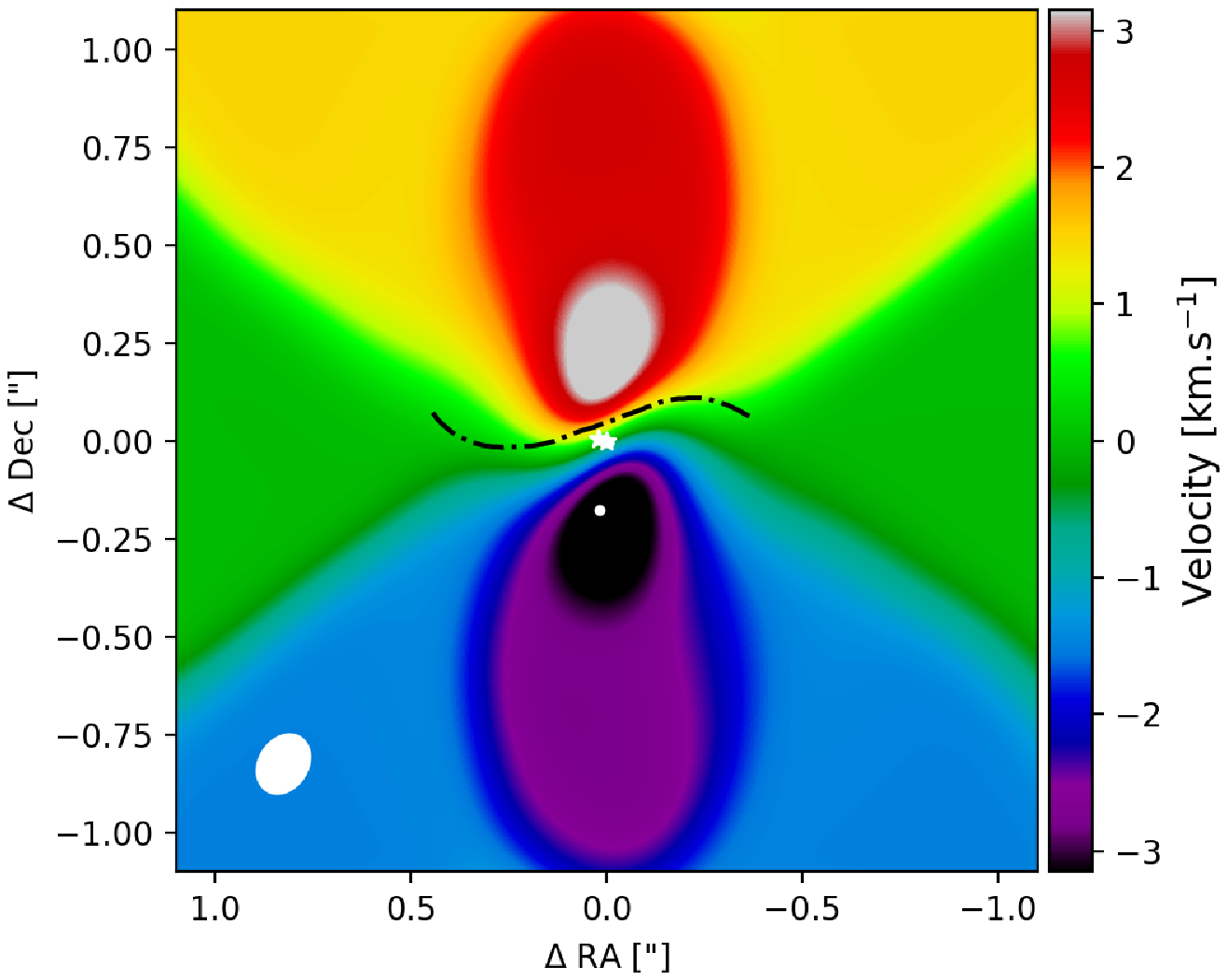}
  \caption{ The comparison between the $^{12}$CO $J=1-2$ moment 1 maps of the observations from {\protect\cite{Bi2020}} (upper panel) and our two hydrodynamical simulations (lower-left panel, run3, and  lower-right panel, run4). The observation and synthetic images are performed with a beam of 0.122" $\times$ 0.159" with a position angle of $-32.3\degree$ (bottom left corner). The dotted-dashed line highlights the shape of the twist. The velocities in the upper and lower panel are shown with respect to the rest velocity of 13.5km/s. We plot the binary (white stars) and planet (white dot) in the synthetic images. There are no localised  artefacts surrounding the planet.  We have copied the twist line in the observation panel and displayed it on our synthetic images.  The twist in the inner regions of the synthetic image with a planet is more consistent with the observation.}
\label{fig::CO_kinematics}
\end{figure*}

 We further examine the evolution of the disc with $H/r = 0.05$  (run4) at times $t = 0\,, 100\,, 180\,, 230\, \rm P_{planet}$ in Fig~\ref{fig::disc_params_planet}. There is initially a pre-carved gap in the surface density profile, shown by the trough in the curve centered on $100\, \rm au$. Since we start with an initial gap, the disc is essentially broken and this break propagates outwards to about $150\, \rm au$ at a time $t = 100\, \rm P_{\rm planet}$. At  $t = 180\, \rm P_{\rm planet}$ the break has propagated to the outer regions of the disc and slowly dissipates. After $t = 180\, \rm P_{\rm planet}$ the initial break has been fully dissipated, however, the $1\, \rm M_{\rm J}$ planet is close to coplanar to the disc and it begins to open a new gap which is shown by the dip in the surface density profile at $t = 230\, \rm P_{\rm planet}$. During the simulation, the planet migrates inward to $\sim 75\, \rm au$ by a time of $t = 230\, \rm P_{\rm planet}$. The center of the break is located at $\sim 75\, \rm au$.  The peaks in the eccentricity profile correspond to the break location. At $t = 230\, \rm P_{\rm planet}$ the inner regions of the disc have a larger eccentricity than the outer parts of the disc. From observations, the inner ring is more eccentric that the middle and outer rings, with an estimated eccentricity of $\sim 0.2$ \citep{Bi2020,Kraus2020}.  Moreover, Fig.~\ref{fig::planet_mass} shows the planet mass as a function of time for an initially $1\, \rm M_{\rm jup}$ planet. The planet mass increases significantly after $t = 200\, \rm P_{planet}$, which corresponds to the time the planet has realigned with the disc.

 \begin{figure*}
 \centering
  \includegraphics[width=0.5\columnwidth]{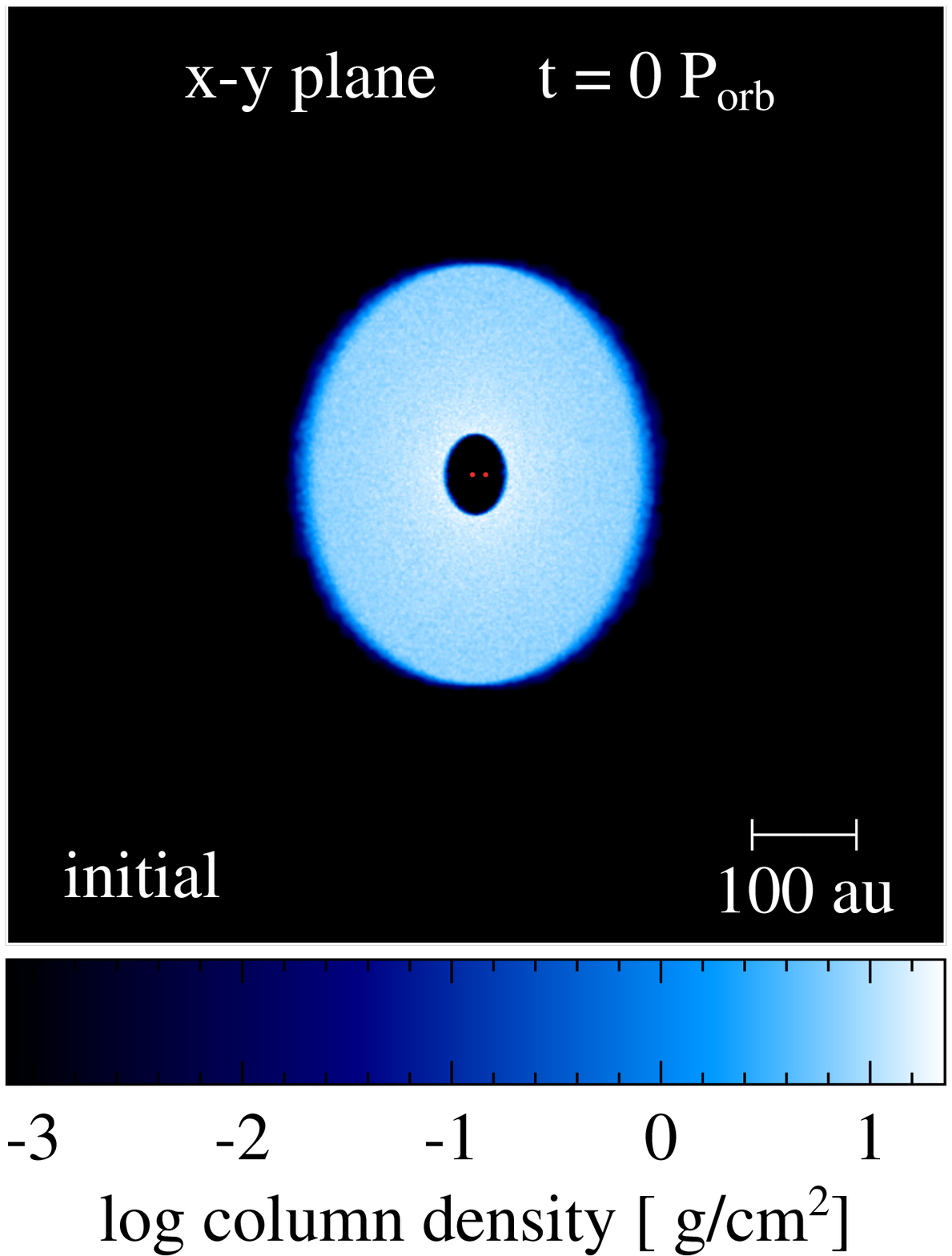}
    \includegraphics[width=0.5\columnwidth]{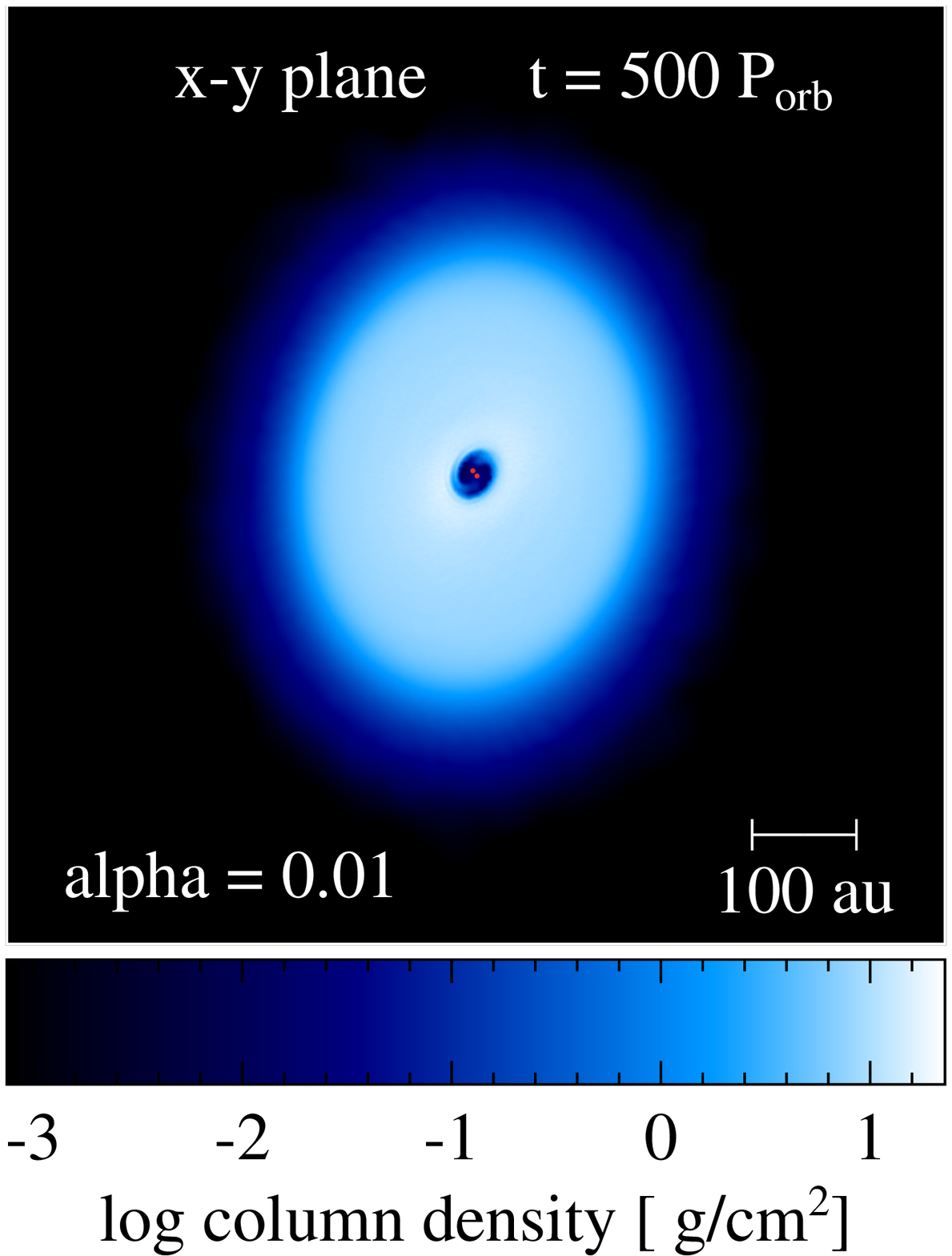}
      \includegraphics[width=0.5\columnwidth]{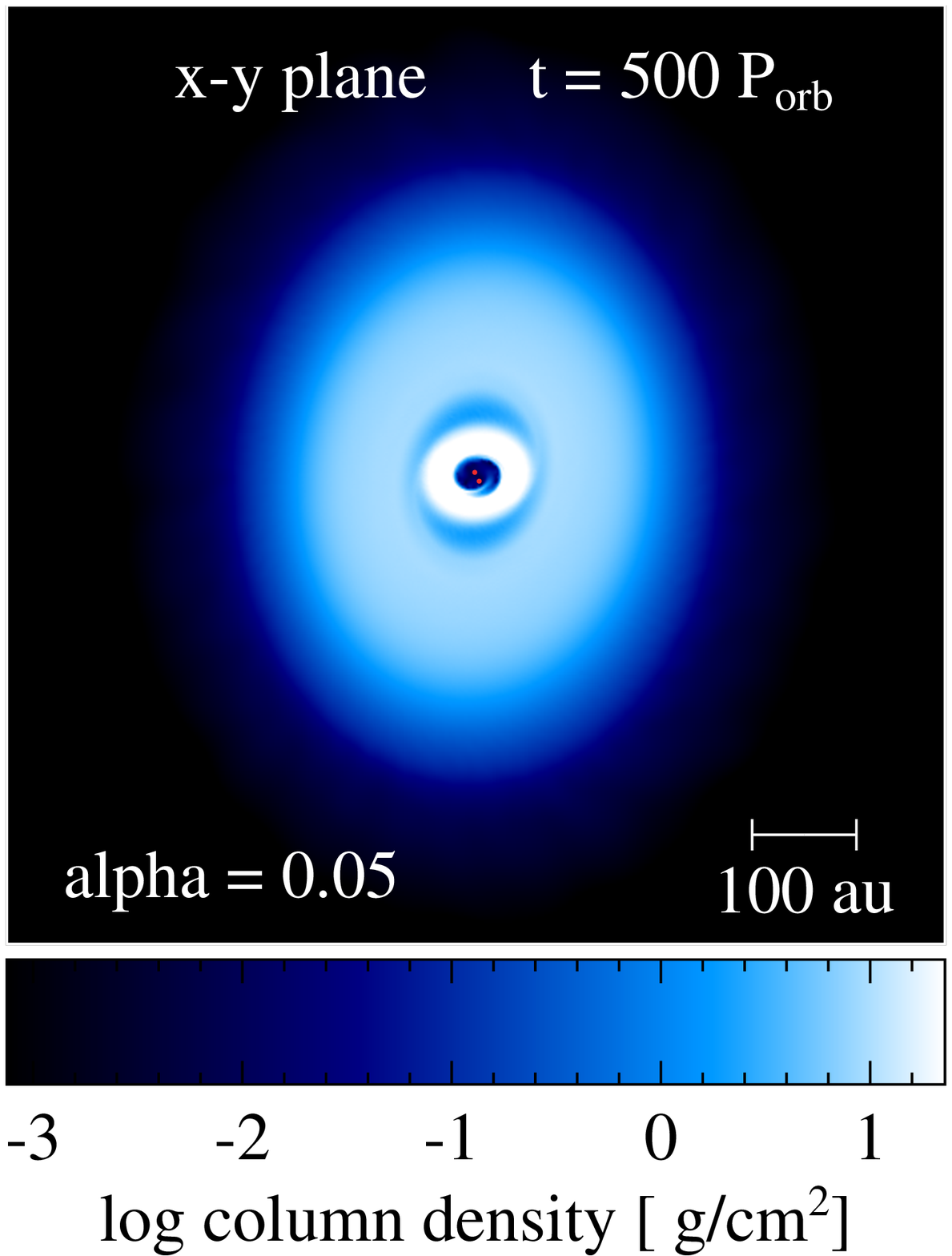}
        \includegraphics[width=0.5\columnwidth]{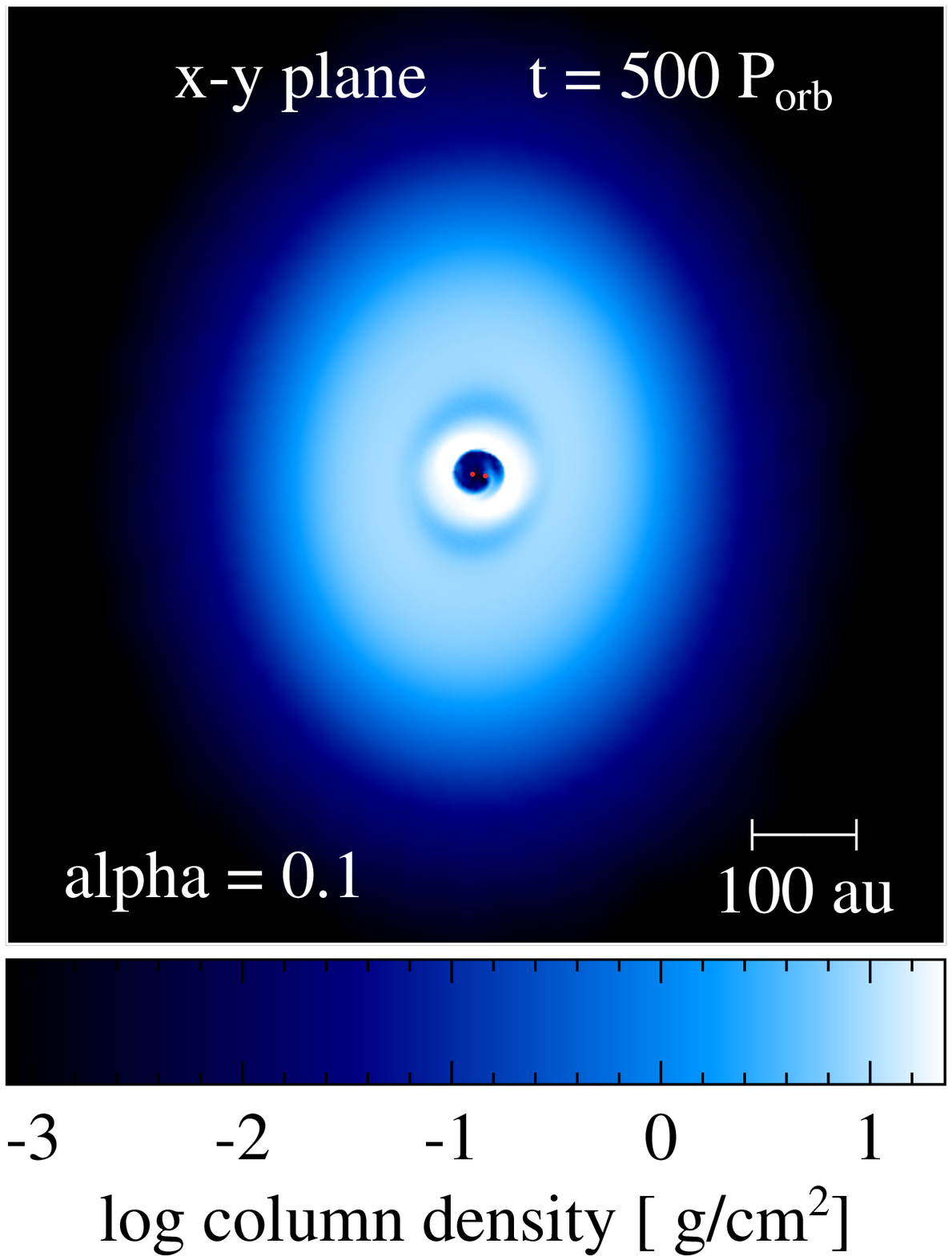}
  \caption{Disc evolution for a circumbinary disc  with varying the $\alpha$--viscosity parameter. The binary components are shown by the red dots. Beginning from the left-most panel we show the initial conditions,  $\alpha = 0.01$ (run6 from Table~\ref{table::setup}),  $\alpha = 0.05$ (run7), and then $\alpha = 0.1$ (run8). 
  The disc evolution is shown at a time $t = 500\, \rm P_{orb}$. The colour bar denotes the gas density. We show the view looking down on to the binary orbital plane, the $x$--$y$ plane. For higher viscosity values, the disc is more prone to breaking.}
\label{fig::alpha_splash}
\end{figure*}

 \begin{figure}
 \centering
  \includegraphics[width=\columnwidth]{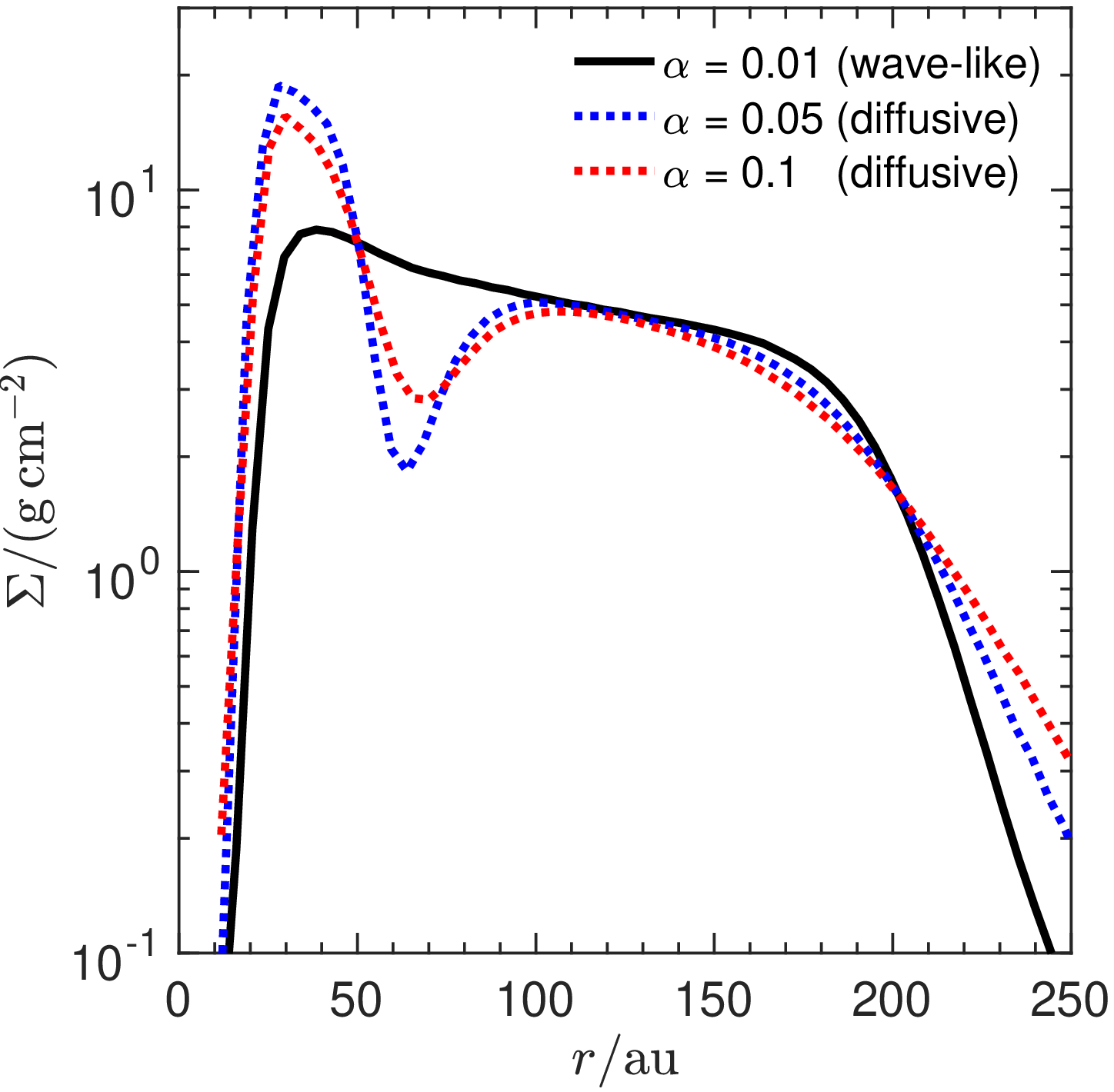}
  \caption{ The surface density profile taken from run6 ($\alpha = 0.01$, black),  run7 ($\alpha = 0.05$, blue) and run8 ($\alpha = 0.1$, red) at a time $t = 500\, \rm P_{orb}$. The wave-like regime is shown by the solid line and the diffusive regime is denoted by the dotted lines. At lower viscosities more typical of protoplanetary discs, the smooth surface density profile shows no sign of breaking.}
\label{fig::alpha}
\end{figure}

\subsection{CO kinematics}
As we have recovered the disc structure inferred by \citet{Bi2020}, we now consider the synthetic CO moment 1. Figure~\ref{fig::CO_kinematics} shows the comparison between the $^{12}$CO $J=1-2$ moment 1 maps of our simulations. The upper panel reproduces the observation from \citet{Bi2020} for convenience. For a regular Keplerian disc, the green regions should represent a well-defined ``butterfly-like'' pattern. However, the observations show a twisted pattern inside $\sim0.2"$, which can be accounted for by a warp in the disc where there is a misalignment between the inner and outer portions. The twist is outlined in the insert. This twisted pattern has also been seen in the discs around HD 142527 \citep{Casassus2015,Marino2015}, HD 143006 \citep{Benisty2018,Perez2018},  HD 97048 \citep{VanDerPlas2017}, AB Aurigae \citep{Poblete2020}, IRS 48 \citep{Calcino2019}, and MWC 758 \citep{Calcino2020}. The lower-left panel shows the CO velocity map for the simulation of only a circumbinary disc (run 3 from Table~\ref{table::setup}) and  the lower-right panel in Fig.~\ref{fig::CO_kinematics} shows the CO velocity map for our simulation that includes a planet (run 4).

In both synthetic images, the inner portions of the map, $\sim0.2"$, shows a twist in inner regions which is consistent with the misaligned inner binary.  While the broad features in both maps from the simulations are roughly in agreement with the features identified in the observations, only the simulation with the planet additionally has a matching disc structure.



%

\subsection{Comparison with \citet{Kraus2020}}
 Up to this point we have shown, from our SPH simulations modeling a circumtriple and circumbinary disc, that the disc does not break due to the binary or triple star system. This motivated our previous section, where we invoked a planet to break the disc at the required radius. However, \cite{Kraus2020}  conducted SPH simulations and found that the torque from the triple system can effectively break the disc. Here we explore the differences between our model and the model from \cite{Kraus2020}. 

\citet{Kraus2020} also conduct their simulation using gas particles in a  Lagrangian SPH code \citep{Bate1995,Price2007}. They use fewer particles over a smaller radial extent, with $8\times 10^5$ between $20\, \rm au$ and $200\, \rm au$ where we use $1\times 10^6$ particles between $40\, \rm au$ and $400\, \rm au$. Their disc surface density profile  is shallower than ours with $\Sigma \propto r^{-1/2}$. Both of our models use a fixed disc aspect ratio of $H/r = 0.05$. One  difference of their setup compared to this work is that their disc is initially set up orbiting a single mass of $5.26\, \rm M_{\odot}$, which is the mass of the total triple star system. The disc is then simulated with this single gravitational mass so that any transient features due to the initial conditions are dissipated. Once the disc reaches a steady-state, the central mass is replaced with the three stars and the disc is reoriented on the center of mass of the system and inclined by $38\degree$ relative to the plane of the (AB) -- C orbit. Along with this, any material within $40\, \rm au$ is removed. They also neglect disc self-gravity and the gravitational effect from the disc onto the stars is ignored. The results of their simulation showed an inner ring break off from the outer disc and precess independently. Moreover, the eccentricity measured in the inner ring of their simulation is $\sim 0.15$, the eccentricity measured from observations of the inner most ring of $\sim0.2$.

Since we have provided adequate evidence that the binary setup and parameters do not cause the discrepancy in the disc evolution between the two models, we now explore the effects of varying the disc parameters. The simulations discussed here are runs 6-10 from Table~\ref{table::setup}. Here, we repeat the simulation setup from \cite{Kraus2020} but we use a binary system rather than a triple system (we tested the triple system back in Section~\ref{sec:triple}) and do not clear particles within $40\, \rm au$. We explore how sensitive the $\alpha$ parameter is to induce disc breaking. Furthermore,   \cite{Kraus2020} initially clear any particles within $40\, \rm au$, however, we also investigate how tuned the disc breaking is to the initial inner disc radius. Our effort in exploring how these parameters contribute to disc breaking will resolve the discrepancies in the competing models of \cite{Bi2020} and \cite{Kraus2020}.

 \begin{figure}
 \centering
  \includegraphics[width=\columnwidth]{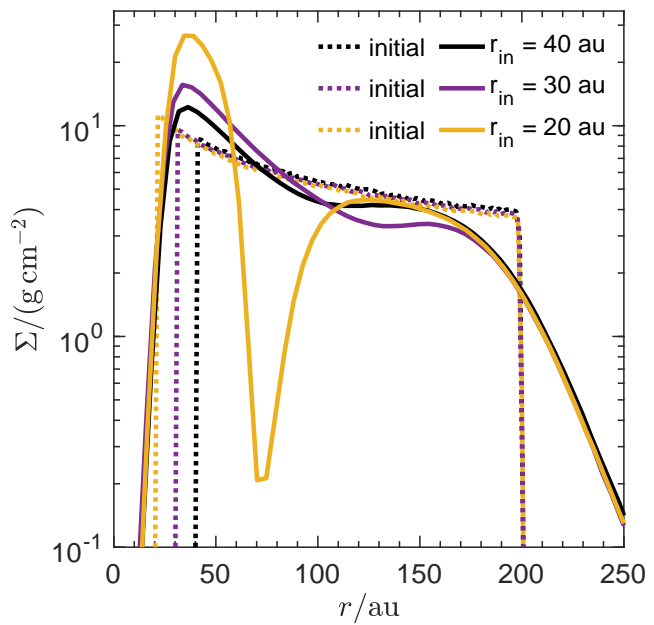}
  \caption{ The surface density profile taken from  run6 ($r_{\rm in} = 40\, \rm au$, black), run9 ($r_{\rm in} = 30\, \rm au$, purple) and  run10 ($r_{\rm in} = 20\, \rm au$, yellow) at a time $t = 1000\, \rm P_{orb}$. Disc material close to the binary will cause the disc to break, showing a deep depression in the surface density profile.   }
\label{fig::rin}
\end{figure}

 \begin{figure*}
 \centering
  \includegraphics[width=\columnwidth]{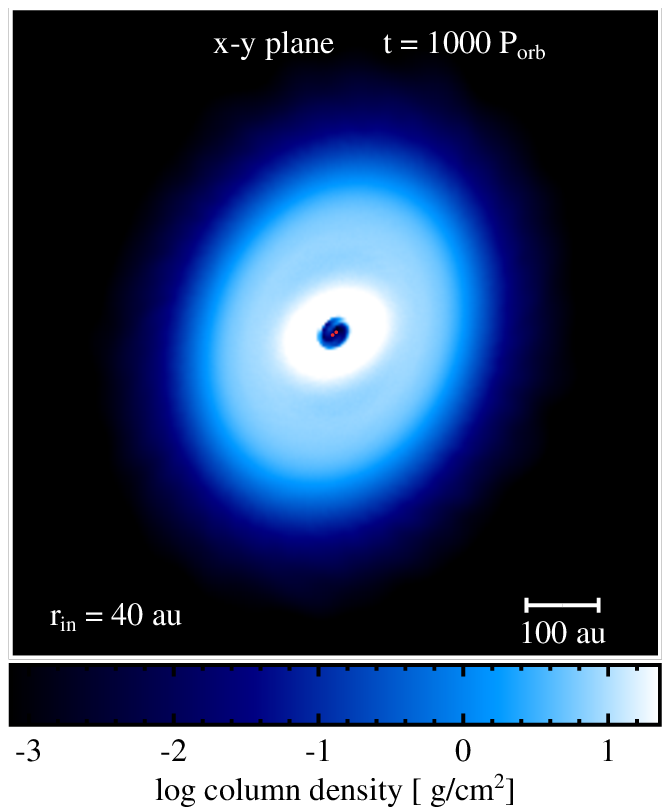}
  \includegraphics[width=\columnwidth]{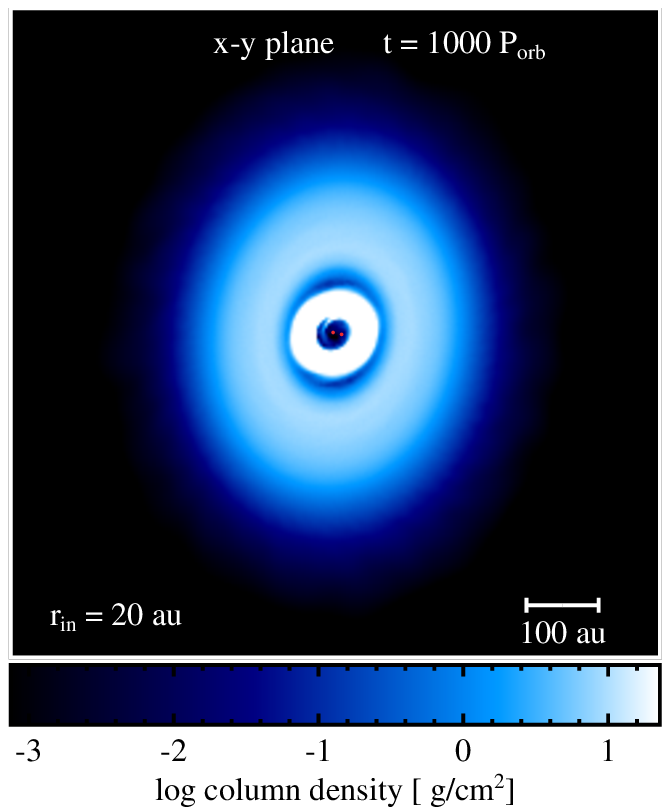}
  \caption{Disc evolution for a circumbinary disc  with varying the the initial inner disc radius. We show $r_{\rm in} = 40\, \rm au$ ( run6, left panel) and $r_{\rm in} = 20\, \rm au$ ( run10, right panel). The binary components are shown by the red dots. The disc evolution is shown at a time $t = 1000\, \rm P_{orb}$. The colour bar denotes the gas density. We show the view looking down on to the binary orbital plane, the $x$--$y$ plane. Disc breaking occurs when there is too much material initially close to the binary.}
\label{fig::rin_splash}
\end{figure*}

\subsubsection{Viscosity}

To better visualize the breaking when varying $\alpha$, we show the disc structure in Fig.~\ref{fig::alpha_splash}. Beginning from the left-most panel, we show the initial condition, followed by the disc structure at a time $t=500\,P_{\rm orb}$ for the simulations with  $\alpha = 0.01$ (run6 from Table~\ref{table::setup}), $\alpha = 0.05$ (run7), and $\alpha = 0.1$ (run8). The disc is initially misaligned by $38\degree$ and we view each disc in the $x$--$y$ plane. The two discs that are in the diffusive regime (the two right-most panels) show the disc breaking. The disc that is in the bending wave regime,  $\alpha = 0.01$, has no signs of disc breaking. In the wave-like regime the communication throughout the disc is rapid 
and allows the disc to maintain a coherent disc-like structure while the diffusive disc cases break. The breaking criteria in hydrodynamical simulations is also dependent on resolution, however the number of initial particles used here provides adequate resolution \cite[comparing the values in our Table~\ref{table::setup} with Fig. 8 in][]{Nealon2015}.

Figure~\ref{fig::alpha} shows the surface density profile for discs with three different $\alpha$--viscosity parameters.  The lower viscosity value provides a smooth surface density profile which means that the disc has not broken. For $\alpha = 0.05,\, 0.1$ (diffusive-type discs), there is a  dip in the surface density profile which means that the disc is broken. For the lower $\alpha$ and $H/r$ expected in GW Ori, Fig.~\ref{fig::alpha_splash} confirms that although warping is expected, breaking is not.

In the viscous regime, the dominant torques are the viscous torque and the precession torque. Therefore, for high $\alpha$ values the viscous torque dominates the precession torque, meaning that the disc is less likely to break \cite[e.g.,][]{Nixon2013a}. For $\alpha = 0.05$ the disc breaks more easily than when $\alpha = 0.1$. The dip in the surface density is more prominent in the $\alpha=0.05$ case than when $\alpha=0.1$ because, in the diffusive regime, for high $\alpha$ values the disc becomes comparatively harder to break \citep{Nixon2013a}.  

In the wave-like regime, the breaking criteria depends upon the global precession rate compared to the communication timescale (as described in Section 3.1). Neither of these timescales depend directly upon $\alpha$ (see equations 3 and 8). However, as shown in Figure~\ref{fig::alpha},
 the inner edge of the disc is further out for lower $\alpha$ and for larger $\alpha$ values the inner edge can live closer to the binary.  The global precession timescale is sensitive to the value of the disc inner radius. The precession timescale increases with $r_{\rm in}$. Thus, in the wave-like regime, for lower $\alpha$, $r_{\rm in}$ increases, the precession timescale therefore increases and the disc is less likely to break. Since $\alpha$ in the GW Ori disc may be even lower than the 0.01 value we have considered, the disc may be even less susceptible to breaking that in the simulation shown here. 

\subsubsection{Initial location of the inner radius}
Figure~\ref{fig::rin} shows the surface density profiles of  three discs with initial inner radii of $40\, \rm au$ (black line,  run6 from Table~\ref{table::setup}),  $30\, \rm au$ (purple line, run9), and $20\, \rm au$ (yellow line,  run10). The surface density profile of the larger inner radius simulation remains smooth which indicates that the disc is not broken. This disc structure is the same for when $r_{\rm in} = 30\, \rm au$. Meanwhile, the disc with a smaller initial inner radius, $20\, \rm au$, has a dip in the surface density profile. Figure~\ref{fig::rin_splash} shows the disc structure given two different initial radii ($40\, \rm au$ and $ 20\, \rm au$). A clear break can be seen for when $r_{\rm in} = 20\, \rm au$, while the disc structure with $r_{\rm in} = 40\, \rm au$ remains smooth. Whether the disc breaks or not thus depends sensitively on the inner radius that the disc is initialised with.

\section{Discussion}
\label{sec::discussion}

\subsection{Growth of an inclined planet}
 The results of the simulations described above shows that if a planet forms in a misaligned disc and is massive enough to carve a gap, it can lead to an effectively broken disc. The inner disc precesses faster than the viscous spreading (shown in the third panel in Fig.~\ref{fig::disc_params_planet}) and so the disc parts remain misaligned, as seen in the second panel. This process may repeat itself -- if the planet becomes misaligned again, the break will propagate outward with the warp and dissipate until the planet's tilt oscillates back to a coplanar orientation with respect to the disc. The planet will then carve another gap, and so on. Each time the planet becomes aligned with the disc, the mass of the planet increases. From Fig.~\ref{fig::planet_mass}, the planet mass increases significantly after $t = 200\, \rm P_{planet}$, which corresponds to the time the planet has realigned with the disc. This implies that planets formed in a misaligned disc may become more massive than planets formed within a coplanar disc if they are able to carve multiple gaps in the disc due to their evolution where they become misaligned to the disc and later realigned. 

The proposed inclined planet in GW Ori may be difficult to observe. Planets are large separations, regardless of their planetary radius, are more difficult to detect than giant planets at small separations.
Giant planet detections are more common around A stars and that wide-orbit planets are more conducive around high-mass stars \citep{Johnson2010,Reffert2014}. Moreover, the occurrence rate of directly imaged giant planets is on the order of $10\%$ \citep{Galicher2016,Meshkat2017,Nielsen2019,Baron2019}. This being said, the results displayed in the CO kinematics in Fig.~\ref{fig::CO_kinematics} show no localised artefacts surrounding the giant planet, which suggest that detection would be challenging. 

Furthermore, \cite{Kraus2020} presented SPHERE and GPI coronagraphic-polarimetric observations of GW Ori. They are best suited to reveal disc structures by exploiting the fact that direct starlight is not polarized but scattered light from the disc is. These observations are not ideal for searching for thermal emission from faint companions next to bright stars. High contrast imaging in total intensity employing various kinds of speckle suppression techniques \cite[e.g., ADI,][]{Marois2006} are needed for this purpose. In addition, searching for companions in discs faces the difficulty that disc signatures may compromise point source recoveries \citep{Maire2017}. So far, the only wildly accepted planet detections in disks are PDS 70b,c \citep{Keppler2018,Haffert2019}. The small stellocentric distance of the predicted gap opening planet in our model ($\sim0.25''$) also poses a challenge. Additional challenges include the multiplicity of the central stars (potentially complicating coronagraph deployment) and their high luminosities \cite[bolometric luminosity ~ 50 $L_\odot$,][]{Fang2014} that results in large flux ratio between the stars and planets.

\subsection{Connecting the disc breaking and the dust structures}

In this work we exclusively simulated a gas disc. However, the observations by \citet{Bi2020} are of the dust in the gas disc. Dust particles undergo various degrees of coupling depending on their Stokes number \cite[e.g.][]{Birnstiel2010}. Initially well coupled dust grains grow, over time, to higher Stokes number and gradually decouple from the gas disc. If significant decoupling occurs in a misaligned disc, the dust particle orbits may evolve independently of the gas disc due to differential precession and the dust structure will not maintain its coherent structure \cite[e.g.,][]{Nesvold2016,Aly2020}.  As the disc rings in GW Ori are observed as coherent structures, the dust must be well coupled to the gas, justifying our use of gas-only simulations to infer the observed structures.  We assume that there are two distinct rings in the GW Ori system, ring 1 and ring 23. The individual rings 2 and 3 have similar inclination and phase angle which suggest that they are not broken but are instead mildly warped. An additional planet located at $100\, \rm au$ is able to explain the misalignment between rings 1 and 23. Such a low-mass planet may be able to explain the warping in the outer ring 23 but we note there are also alternative scenarios that do not require planets that could explain the separation between the outer dust rings \cite[e.g.,][]{Flock2015,Dullemond2018,Suriano2018,Suriano2019,Riols2019,Tominage2020}.

\subsection{Viscosity}
\label{visc}

In Section 5.4.1 we showed that our conclusions rely on a robust estimate for the viscosity in the disc (parameterised by $\alpha$) and the aspect ratio ($H/r$). The viscosity is the fundamental process on how accretion discs evolve \citep{Lynden-Bell1974,Pringle1981}. It determines the transport of mass and angular momentum within the disc, which in turn provides how much energy is released. Viscosity transports angular momentum outwards, allowing matter to spiral inwards in a disc. The disc viscosity parameter, $\alpha$, can be  estimated from observations. The simple estimate of $\alpha$ comes from comparing estimated disc masses, $M_{\rm d}$, with estimated central accretion rates, $\dot{M}_{\rm c}$, and from these deducing an accretion timescale $\tau_{\nu} \sim M_{\rm d}/\dot{M}_{\rm c}$ \citep[e.g.][]{Lodato2017,Martin2019}. By measuring the sound speed $c_{\rm s}$ and the disc height $H$ in the outer disc, the viscosity value can be determined \citep{Hartmann1998}.

 In the past, the upper limit for the $\alpha$ value for protostellar discs was estimated to be $\alpha \sim 0.01$ on distance scales $10$--$100\, \rm au$ \citep{Hartmann1998,Hartmann2000,Trapman2020}, but recent observations show that the upper limit is closer to $\alpha \sim 0.001$. \cite{Andrews2009} observed protoplanetary discs in Ophiuchus and found $\alpha \sim 0.0005-0.08$ for radius $R = 10\, \rm au$. \cite{Hueso2005} found that $0.001 < \alpha < 0.1$ for DM Tau and $4\times 10^{-4} < \alpha < 0.04$ for GM Tau. More recently, \cite{Rafikov2016} used a self-similar disc solution to calculate $0.0001 < \alpha < 0.04$ for resolved disc observations by ALMA. \cite{Ansdell2018} then refined these calculations by measuring the gas disc size and found $0.0003 < \alpha < 0.09$. \cite{Pinte2016} measured the dust scale height in HL Tau, and estimated the turbulent viscosity coefficient to be a few $10^{-4}$. The turbulence levels in discs using ALMA gas observations had been directly measured giving $\alpha \sim 0.001$ \cite[e.g.,][]{Flaherty2015,Flaherty2017,Teague2018}. Moreover, there is evidence that the characteristic ``double gaps'' in HL Tau, TW Hya, and HD 169142 are produced by a low mass planet, which requires a low disc viscosity \cite[e.g.,][]{Dong2017,Dong2018b}. From observations of  protoplanetary discs, the disc is certainly expected to be in the bending-wave regime rather than the viscous regime.  In the context of Sections~\ref{sec:triple},~\ref{sec::woutaplanet} and Fig.~\ref{fig::alpha}, our results confirm  that the gaps in the circumtriple disc around GW Ori are not produced by the binary (or triple star) torque.

\subsection{Inner disc radius}
 We simulated three different initial disc radii, $40\,, 30\,, 20\,, \rm au$ and showed that the binary torque is able to break the disc for $r_{\rm in} = 20\, \rm au$ but not for $r_{\rm in} = 40\, \rm au$ or $r_{\rm in} = 30\, \rm au$.  Material within our simulations is free to move inwards and so even though it begins at $30$ or $40\, \rm au$, it moves in closer to a radius that is determined by the balance of the tidal and viscous torques \citep[e.g.][]{Lubow2015,Miranda2015,Franchini2019b}.  When the inner radius is smaller, there is much more material in the inner regions of the disc and the precession timescale becomes smaller than the radial communication time-scale \citep{Lubow2018}. The location of the inner radius is thus crucial to determining whether the disc breaks or not and should be considered carefully in future work. The simulation with initial $r_{\rm in}=20\, \rm au$ places more material closer in than there would be in the quasi-steady state. This can only be achieved if the accretion of material is occurring in $r<30$. In the case of GW Ori, observations suggest that the inner radius is located at $\sim32$~au. \cite{Bi2020} adopted an inner disc radius of $32\, \rm au$, which is 3-4 times the AB-C binary semi-major axis, which is also supported by \cite{Czekala2017} and \cite{Kraus2020}.With this larger inner radius, our simulations confirm there should be no breaking.

Formation scenarios suggest that misaligned discs are likely to be formed by misaligned gas falling onto existing binaries or triples \citep{Bate2018}. The large outer radius of GW Ori ($\sim 1300\, \rm au$, \citealt{Bi2020}) and orientation of the outer disc are consistent with this formation mechanism. In this interpretation, as the gas and dust fall inwards, it crosses the region where warping and breaking can occur. As such, we do not predict that material should be found inside the warping radius with the observed orientation of the outer disc. Although not the focus of this study, this is highly relevant to the inner radius used in the initial conditions, which is used here and shared with \citet{Kraus2020}.


\subsection{Consideration of planets}

A possible mechanism to produce gaps and misalignment in the GW Ori disc may be due to the presence of planets. High-mass planets exert a tidal torque that overpowers the local viscous torque and form a gap in the gas \citep{Papaloizou1984,Bryden1999}. In Section~\ref{sec::planet}, we showed that a giant planet that forms initially coplanar to the disc can produce a strong warp, if the disc is sufficiently thin. Therefore, the misalignment between the inner and middle rings in GW Ori may be caused by the presence of planets. However, if the disc aspect ratio is larger, the presence of the dust gaps in GW Ori must be produced by a low mass planet that is well coupled to the gas disc. The low mass planet will be well coupled to the gas forcing it to precess at the same rate as the disc. \cite{Dipierro2016} ran SPH simulations of a gas and dust disc with an embedded low-mass planet. The planet is effective in opening a gap in the dust but not in the gas. However, this would not explain the observed misalignment between the inner and middle rings since the gas disc would maintain a flat coherent structure.

If misaligned planets are present around the hierarchical triple star system, they would be difficult to detect. The torque produced by the binary affects the formation processes of planets embedded in the circumbinary gas disc compared to discs around single star systems \citep{Martin2014,Fu2015a,Fu2015b,Fu2017}. When a giant planet is formed within a misaligned disc, the torque from the binary  prevents the planet from remaining coplanar to the binary orbital plane \citep{Lubow2016,Martin2016,Pierens2018}. The probability of detecting inclined planets through the transit method is lower than coplanar planets. Follow-up observations have revealed that $\sim 2.5\%$ of planets are in triple and multiple systems \citep{Roell2012,Fragione2019}. However, no planet in a circumtriple orbit has been detected.  If a planet (or planets) are the cause of the dust gaps in the circumtriple disc around GW Ori, then they would be the first circumtriple planet(s).

\section{Conclusions}
\label{sec::conclusions}

We have examined the origin of the coherent dust structures around the GW Ori hierarchical triple system.  \cite{Bi2020} first suggested that the break cannot be caused by the torque from the observed triple star system. More recently, \cite{Kraus2020}  conducted SPH simulations and stated that the triple star torque was the cause for the break.  In this work, we carried out extensive SPH simulations to explore the discrepancy between the two models.

 First, we compared through 3-dimensional hydrodynamical simulations that a circumbinary disc evolves in a similar fashion as a circumtriple disc in the context of the GW Ori system. Second, we tested the differences in the binary parameters between the two models and found that the disc evolves in a similar fashion, independently of the binary parameters used. We then examined the disc viscosity and the inner radius of the disc since these two parameters heavily impact the criteria for disc breaking. We found that when $\alpha = 0.01$, the disc is strongly warped but does not break.  Lastly, we showed a small initial disc radius will cause the disc to break even in the bending-wave regime. However,  this effect is due to the initial conditions of the simulation.  Since $\alpha = 0.01$ is an upper limit for protoplanetary discs \citep{Hartmann1998} and the surface density profile is tapered within the inner regions, our results show that the break in the GW Ori circumtriple disc is not caused by the triple star system. 

We present an alternative scenario to explain the origin of the dust rings in GW Ori, using a  planet (or planets). We find that an initially massive planet can continuously open a gap within a thin disc as the planet's tilt oscillates in and out of the disc plane. For a thicker disc, the viscous spreading is too fast for the planet to maintain the gap. However, a low-mass planet that is well coupled to the gas can still open a gap in the dust \citep[e.g.,][]{Dipierro2016}. In conclusion, we have shown that the break in the GW Ori circumtriple disc is not due to the torque imposed onto the disc by the stars. Therefore, the disc breaking must be caused by undetected planets, which would be the first planets in a circumtriple orbit.

\section*{Data Availability Statement}

The data supporting the plots within this article are available on reasonable request to the corresponding author. A public version of the {\sc phantom} and {\sc splash} codes are available at \url{https://github.com/danieljprice/phantom} and \url{http://users.monash.edu.au/~dprice/splash/download.html}, respectively. {\sc mcfost} is available on request.

\section*{Acknowledgements}

We thank the anonymous referee for helpful suggestions that positively impacted the work. The authors thank Alison Young for providing details of her simulations. The three-body integrations in this work made use of the REBOUND code which can be downloaded freely at http://github.com/hannorein/rebound. Computer support was provided by UNLV's National Supercomputing Center. We acknowledge the use of \textsc{SPLASH} \citep{Price2007} for the rendering of figures \ref{fig::disc_splash}, \ref{fig::planet8_splash}, \ref{fig::alpha_splash} and \ref{fig::rin_splash} as well as \textsc{pymcfost} for \ref{fig::CO_kinematics}. RN acknowledges funding from the European Research Council (ERC) under the European Union's Horizon 2020 research and innovation programme (grant agreement No 681601)  and subsequent support from UKRI/EPSRC through a Stephen Hawking Fellowship (EP/T017287/1). R.D. acknowledges the financial support provided by the Natural Sciences and Engineering Research Council of Canada. C.P. acknowledges funding from the Australian Research Council via FT170100040 and DP180104235.






\bibliographystyle{mnras}
\bibliography{ref} 



\appendix

\section{Supplemental information}
In this appendix we provide additional information on the setup of the hydrodynamical simulations by showing the ranges of the $\alpha$--viscosity parameter, and the shelled-averaged smoothing length per scale height in Table~\ref{table::A1}. 

 \begin{table}
	\centering
	\caption{Additional information for the SPH simulations that lists the  disc aspect ratio at the disc inner and outer edges $H/r$, the minimum and maximum values of the \citet{shakura1973} $\alpha$-viscosity parameter, and the minimum and maximum values of the shelled-averaged smoothing length per scale height $\langle h \rangle/H$.}
	\begin{tabular}{ccccc} 
		\hline
	    Simulation  & $\alpha_{\rm min}$ & $\alpha_{\rm max}$  & $(\langle h \rangle/H)_{\rm min}$ & $(\langle h \rangle/H)_{\rm max}$  \\
		\hline
        \hline
        run0  &  $0.006$ & $0.021$ & $0.21$  & $0.71$ \\
		run1  &  $0.008$ & $0.013$ & $0.27$  & $0.42$ \\
        run2  &  $0.008$ & $0.013$ & $0.27$  & $0.42$ \\
        run3  &  $0.008$ & $0.013$ & $0.17$  & $0.27$ \\
        run4  &  $0.008$ & $0.013$ & $0.27$  & $0.42$ \\
        run5  &  $0.008$ & $0.013$ & $0.17$  & $0.27$ \\
        run6  &  $0.006$ & $0.021$ & $0.21$  & $0.71$ \\
        run7  &  $0.007$ & $0.017$ & $0.21$  & $0.49$ \\
        run8  &  $0.035$ & $0.083$ & $0.21$  & $0.49$ \\
        run9  &  $0.071$ & $0.165$ & $0.21$  & $0.49$ \\
        run10  &  $0.007$ & $0.0184$ & $0.21$  & $0.58$ \\
        \hline
	\end{tabular}
    \label{table::A1}
\end{table}



\bsp	
\label{lastpage}
\end{document}